\documentclass[iop,apj]{emulateapj}  

\usepackage{soul}
\usepackage{datetime}

\def \npreqtwelvepcs {2842~}  
\def \nqtwelvenewkois {1990~}  
\def \nqtwelvenewpcs {855~}  
\def \nvettedkois {2355~}  
\def \nqtwelvevettedpcs {868~}  
\def \nqtwelvevettedfps {1487~}  
\def \nallpcs {3697~}  
\def \nallpcsk {3697}
\def \nallkois {5855~}  

\newcommand{\gcmc}{\ensuremath{\rm g\,cm^{-3}}}

\newcommand{\teff}{\ensuremath{T_{\rm eff}}}

\newcommand{\logg}{\ensuremath{\log{g}}}

\newcommand{\feh}{[Fe/H]}

\newcommand{\rsun}{\ensuremath{R_\sun}}
\newcommand{\msun}{\ensuremath{M_\sun}}

\newcommand{\rstar}{\ensuremath{R_\star}}
\newcommand{\mstar}{\ensuremath{M_\star}}

\newcommand{\rhostar}{\ensuremath{\rho_\star}}
\newcommand{\rhoc}{\ensuremath{\rho_c}}

\newcommand{\rpl}{\ensuremath{R_{\rm p}}}

\newcommand{\rprs}{\ensuremath{R_{\rm p}/R_\star}}

\newcommand{\rearth}{\ensuremath{R_{\earth}}}

\newcommand{\ikt}{{\it Kepler}}
\newcommand{\ik}{{\it Kepler~}}

\newcommand{\oldsys}{2412}
\newcommand{\newsys}{2674}
\newcommand{\oldplans}{3136}
\newcommand{\newplans}{3535}
\newcommand{\oldmulti}{480}
\newcommand{\newmulti}{572}
\newcommand{\oldmultiplans}{1204}
\newcommand{\newmultiplans}{1433}

\newcommand{\newlowkois}{7}
\newcommand{\newhighkois}{655}

\newcommand{\uppers}{96}
\newcommand{\totaluppers}{117}
\newcommand{\downers}{22}

\shorttitle{\ik Catalogue}
\shortauthors{Rowe et al.}
\slugcomment{Revision 1.8 --- \today --- \currenttime}

\begin{document}

\title{Planetary Candidates Observed by \ik V: Planet Sample from Q1-Q12 (36 Months)}

\author{Jason F. Rowe\altaffilmark{1,2,3}, Jeffrey L. Coughlin\altaffilmark{2}, Victoria Antoci\altaffilmark{23}, Thomas Barclay\altaffilmark{3,9}, Natalie M. Batalha\altaffilmark{3}, William J. Borucki\altaffilmark{3}, Christopher J. Burke\altaffilmark{2,3}, Steven T. Bryson\altaffilmark{3}, Douglas A. Caldwell\altaffilmark{2,3}, Jennifer R. Campbell\altaffilmark{17,3}, Joseph H. Catanzarite\altaffilmark{2}, Jessie L. Christiansen\altaffilmark{13},  William Cochran\altaffilmark{4}, Ronald L. Gilliland\altaffilmark{5}, Forrest R. Girouard\altaffilmark{24}, Michael R. Haas\altaffilmark{3}, Krzysztof G. He\l miniak\altaffilmark{6}, Christopher E. Henze\altaffilmark{3}, Kelsey L. Hoffman\altaffilmark{2,3}, Steve B. Howell\altaffilmark{3}, Daniel Huber\altaffilmark{2,3,20}, Roger C. Hunter\altaffilmark{3}, Hannah Jang-Condell\altaffilmark{11}, Jon M. Jenkins\altaffilmark{3}, Todd C. Klaus\altaffilmark{19}, David W. Latham\altaffilmark{25}, Jie Li\altaffilmark{2}, Jack J. Lissauer\altaffilmark{3}, Sean D. McCauliff\altaffilmark{17}, Robert L. Morris\altaffilmark{2}, F. Mullally\altaffilmark{2,3}, Aviv Ofir\altaffilmark{7,8}, Billy Quarles\altaffilmark{3,10}, Elisa Quintana\altaffilmark{3,10}, Anima Sabale\altaffilmark{17}, Shawn Seader\altaffilmark{2,3}, Avi Shporer\altaffilmark{21,22}, Jeffrey C. Smith\altaffilmark{2,3}, Jason H. Steffen\altaffilmark{16}, Martin Still\altaffilmark{3,9}, Peter Tenenbaum\altaffilmark{2,3}, Susan E. Thompson\altaffilmark{2,3}, Joseph D. Twicken\altaffilmark{2}, Christa Van Laerhoven\altaffilmark{14,15}, Angie Wolfgang\altaffilmark{18},  Khadeejah A. Zamudio\altaffilmark{17,3}}

\altaffiltext{1}{Jason.Rowe@nasa.gov}
\altaffiltext{2}{SETI Institute, Mountain View, CA 94043, USA}
\altaffiltext{3}{NASA Ames Research Center, Moffett Field, CA 94035, USA}
\altaffiltext{4}{Department of Astronomy and McDonald Observatory, The University of Texas at Austin}
\altaffiltext{5}{Center for Exoplanets and Habitable Words, The Pennsylvania State University, University Park, PA 16802, USA}
\altaffiltext{6}{Subaru Telescope, National Astronomical Observatory of Japan, 650 North Aohoku Place, Hilo, HI 96720, USA}
\altaffiltext{7}{Weizmann Institute of Science, 234 Herzl St., Rehovot 76100}
\altaffiltext{8}{Institut f\"ur Astrophysik, Universit\"at G\"ottingen, Friedrich-Hund-Platz 1, D-37077 G\"ottingen, Germany}
\altaffiltext{9}{Bay Area Environmental Research Institute, 596 1st Street West, Sonoma, CA 95476, USA}
\altaffiltext{10}{NASA Postdoctoral Fellow}
\altaffiltext{11}{University of Wyoming, Department of Physics \& Astronomy, Laramie, WY 82071}
\altaffiltext{13}{NASA Exoplanet Science Institute, California Institute of Technology, Pasadena, CA, 91106}
\altaffiltext{14}{University of Arizona. 1629
E University Blvd, Tucson AZ}
\altaffiltext{15}{Canadian Institute for Theoretical
Astrophysics, 60 St George St, Toronto, ON M5S 3H8}
\altaffiltext{16}{CIERA --- Northwestern University
2145 Sheridan Road, Evanston, IL 60208}
\altaffiltext{17}{Wyle Laboratories, Moffett Field, CA 94035, USA}
\altaffiltext{18}{University of California, Santa Cruz}
\altaffiltext{19}{Moon Express, Inc., Moffett Field, CA, 94035}
\altaffiltext{20}{Sydney Institute for Astronomy (SIfA), School of Physics, University of Sydney, NSW 2006, Australia}
\altaffiltext{21}{Jet Propulsion Laboratory, California Institute of Technology, 4800 Oak Grove Drive, Pasadena, CA 91109, USA}
\altaffiltext{22}{Sagan Fellow}
\altaffiltext{23}{Stellar Astrophysics Centre, Aarhus University, Ny Munkegade 120, DK-8000 Aarhus C, Denmark}
\altaffiltext{24}{Logyx LLC}
\altaffiltext{25}{Harvard-Smithsonian Center for Astrophysics
60 Garden Street, Cambridge, MA 02138}

\begin{abstract}

The \ik mission discovered \npreqtwelvepcs exoplanet candidates with 2 years of data. We provide updates to the \ik planet candidate sample based upon 3 years (Q1-Q12) of data. Through a series of tests to exclude false-positives, primarily caused by eclipsing binary stars and instrumental systematics, \nqtwelvenewpcs additional planetary candidates have been discovered, bringing the total number known to \nallpcsk.   We provide revised transit parameters and accompanying posterior distributions based on a Markov Chain Monte Carlo algorithm for the cumulative catalogue of Kepler Objects of Interest.  There are now 130 candidates in the cumulative catalogue that receive less than twice the flux the Earth receives and more than 1100 have a radius less than 1.5 \rearth.  There are now a dozen candidates meeting both criteria, roughly doubling the number of candidate Earth analogs.  A majority of planetary candidates have a high probability of being bonafide planets, however, there are populations of likely false-positives. We discuss and suggest additional cuts that can be easily applied to the catalogue to produce a set of planetary candidates with good fidelity.  The full catalogue is publicly available at the NASA Exoplanet Archive.

\end{abstract}

\keywords{planetary systems --- planets: KOIs --- Facilities: \facility{The \ik Mission}.}

\section{Introduction}
\label{intro} 

The \ik instrument is a 0.95 meter aperture, optical (420 - 915 nm), space-based telescope that employed 42 CCDs to constantly observe 170,000 stars over a field of view (FOV) of 115 square degrees \citep{Koch2010} with a combined noise on 12th magnitude solar-type stars (intrinsic and instrument) of 30 ppm \citep{Gilliland2011} on a 6-hour time-scale. \ik searches for the periodic drops in brightness which occur when planets transit their host star, thusly seeking to identify new extrasolar planets. The primary objective of the \ik Mission is to determine the frequency of Earth-like planets around Solar-like stars \citep{Borucki2010a}.


A series of previously published \ik catalogue papers presented an increasingly larger number of planet candidate discoveries as additional observations were taken by the spacecraft \citep{Borucki2011a,Borucki2011b,Batalha2013,Burke2014}. These catalogues have been used extensively in the investigation of planetary occurance rates \citep[e.g.,][]{Youdin2011,Howard2012,Dressing2013,Fressin2013,Dong2013,Mulders2014,Foreman-Mackey2014}, determination of exoplanet atmospheric properties \citep[e.g.,][]{Coughlin2012,Esteves2013,Demory2014,Sheets2014}, and development of planetary confirmation techniques via supplemental analysis and follow-up observations \citep[e.g.,][]{Moorhead2011,Morton2011,Steffen2012,Ford2012,Fabrycky2012,Santerne2012,Adams2012,Colon2012,Adams2013,Barrado2013,Law2014,Lillo-Box2014,Muirhead2014,Plavchan2014,Rowe2014,Dressing2014,Everett2014}. Furthermore, systems identified as not-planetary in nature have yielded valuable new science on stellar binaries, including eclipsing \citep[e.g.,][]{Prsa2011,Slawson2011,Coughlin2011}, self-lensing \citep{Kruse2014}, and tidally interacting systems \citep[e.g.,][]{Thompson2012}
This paper uses 3 years (Quarters 1--12; Q1--Q12) of \ik photometry to search for new planet candidates, thus enabling for the first time the detection of Earth-like exoplanets that have periods around one year (given that a minimum of three transits are needed for detection.) With this increased sensitivity also comes setbacks --- the instrument is sensitive to a significant number of false positives at periods close to one year due to the spacecraft's heliocentric orbit, combined with a 90 degree boresight rotation every $\sim$\,90 days and electronic, rolling band systematics present in a few CCD modules. Additionally, the number of false positives due to contamination increases with increased sensitivity, as variable stars can induce low-amplitude false positives signatures in sources up to tens of arcseconds away \citep{Coughlin2014}.

In this work we present new methods to eliminate these false positives and introduce a streamlined planet vetting procedure and product set. As a result, we designate an additional \nqtwelvenewpcs planet candidates to bring the cumulative total of \ik planet candidates to \nallpcs. We also present the uniform modeling of all transiting planet candidates utilizing a Markov Chain Monte Carlo (MCMC) algorithm that provides robust estimates of the uncertainities for all of the planet parameters. The posterior distributions allow us to study the planet population in detail and assess the reliability of the most Earth-like candidates.

\section{Detection of Transit-Like Signals}

\subsection{Q1-Q12 Threshold Crossing Events}

\label{tcesec}

We began with the transit-event candidate list from \citet{Tenenbaum2013} based on a wavelet, adaptive matched filter to search 192,313 \ik targets for periodic drops in flux indicative of a transiting planet. Detections are known as Threshold Crossing Events (TCEs). Tenenbaum et al. utilized three years of \ik photometric observations (Q1--Q12) --- the same data span employed by this study based on SOC 8.3 as part of Data Release 21 \citep{Thompson2013}. The authors found a total of 18,406 TCEs on 11,087 individual stars that passed a number of initial diagnostic criteria, such as having a Multiple Event Statistic (MES --- a measure of signal-to-noise) greater than 7.1, having at least 3 transits, and passing some basic false positive tests. For more information, see \citet{Tenenbaum2013}. It should be noted that eclipsing binary candidates identified by the Kepler Eclipsing Binary Working Group (EBWG)\footnote{http://keplerebs.villanova.edu} at the time were excluded from this transit search.

\begin{figure}[h]
\centering
\begin{tabular}{c}
\includegraphics[width=\linewidth]{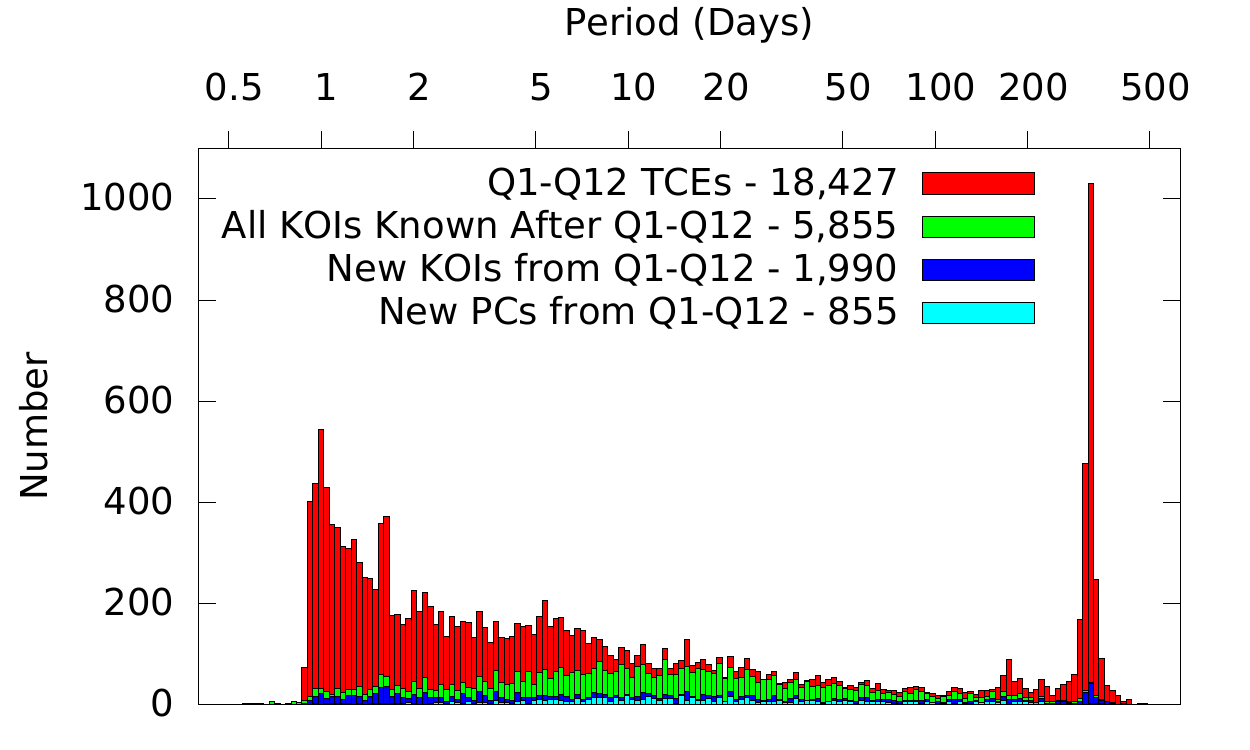} \\
\includegraphics[width=\linewidth]{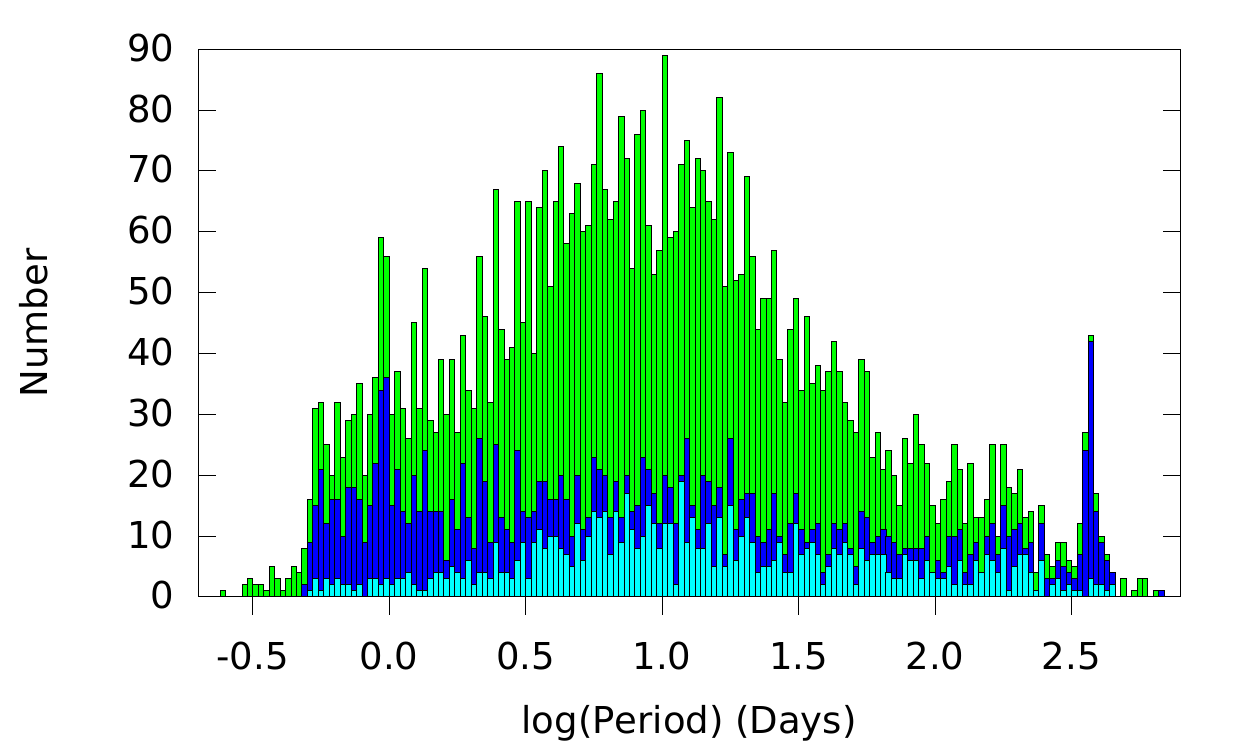} 
\end{tabular}
\caption{Period histogram for various populations. All Q1-Q12 TCEs from \citet{Tenenbaum2013} are shown in red in the top panel. All existing KOIs, after completion of Q1-Q12 TCERT vetting, are shown in green. The new KOIs created as a result of the Q1-Q12 TCERT activity are shown in blue. Finally, the new Planet Candidates (PCs) designated due to the Q1-Q12 TCERT activity are shown in cyan. The top panel shows the full vertical range, while the bottom panel shows a limited vertical range with only KOIs plotted.}
\label{tcefig}
\end{figure}

In Figure~\ref{tcefig} we plot a histogram of the period distribution of all 18,406 TCEs in red. The distribution of transiting exoplanets corrected for geometric effects and signal-to-noise ratio (S/N) as a function of period has been observed to be relatively flat in log space \citep{Howard2012,Fressin2013}. As can be seen in Figure~\ref{tcefig}, there is a large excess in the number of TCEs at both short periods ($\lesssim$10 days) and at long periods ($\sim$372 days). The short-period excess is due mostly to contact binaries and other variable stars that have sinusoidal-like photometric variations on short timescales. The long-period excess is due to stars that fall on CCD modules with significant rolling-band instrumental systematic noise \citep[see][]{VanCleve2009}, which produce sinusoidal-like red noise, once every four quarters. This timescale corresponds to the $\sim$372 day orbital period of the spacecraft. A smaller spike in TCE periods can be seen at $\sim$186 days, where stars fall on CCD modules with rolling band noise every other quarter.

\subsection{Q1-Q10 Threshold Crossing Events}

A run of the \ik pipeline was performed on Q1-Q10 data prior to the run on Q1-Q12, but the results were not published.  Approximately 1000 TCEs were examined that resulted in the production of 360 KOIs with labels 3150 through 3509.  While most of the KOIs generated in the Q1-Q10 run were re-detected in the Q1-Q12 run, $\sim$100 interesting KOIs that appeared to be potential planet candidates were not redetected.  As a result we decided to employ a ``supplemental run" of the DV pipeline to generate Q1-Q12 diagnostics for these Q1-Q10 KOIs. In essence, the DV pipeline was run on Q1-Q12 data for each of these Q1-Q10 KOI targets, with the period and epoch fixed to that found by the Q1-Q10 run. 

\section{Planet Vetting}

Of the 18,406 Q1-Q12 TCEs, four contained data exclusively collected in Q1. These stars were identified as likely evolved stars in Q1 and dropped from the mission target list thereafter, and thus we chose to ignore these TCEs. Of the remaining TCEs, we identified 3,482 that corresponded to previously assigned KOIs via their periods, epochs, and KIC numbers. As we did not desire to re-examine known KOIs, this left 14,920 TCEs that required vetting --- the process whereby some TCEs are designated KOI numbers and then labeled as either Planet Candidates (PCs) or False Positives (FP). Given the large number of TCEs, and that many were known to be due to non-eclipsing variable stars or instrumental systematics (see \S\ref{tcesec}), we decided to employ a two-stage process. The first step, Triage, quickly eliminated obvious false positives so that KOI numbers were assigned only to transit-like TCEs. The second step, Dispositioning, assigned dispositions of either false positive or planet candidate to each TCE desginated as a KOI.

\subsection{Triage}
\label{triagesec}

In Triage, human vetters were given digital documents that contained the {\it Data-Validation} (DV) one-page summary \citep{Wu2010} for each TCE (see \S\ref{dispsec} for more information about the DV one-page summary). On each form, utilizing checkboxes, the human vetters were asked to classify the TCE as belonging to one of four categories: 

\begin{itemize}

\item New Candidate: A TCE that appeared to be possibly due to a transiting or eclipsing astrophysical source, i.e., a transiting planet or an eclipsing binary. 

\item Instrumental: A TCE that was determined to be due to instrumental systematics, such as rolling bands (see \S\ref{tcesec}).

\item Variable Star: A TCE that was deemed to be due to a contact eclipsing binary, pulsating star, spotted star, or any other variable star not associated with a transiting or detached eclipsing source.

\item Low S/N: A TCE that did not appear to have sufficient signal-to-noise to be designated as a KOI. While the formal mission signal-to-noise cutoff is a MES value of 7.1, systematic noise sources can cause the actual signal-to-noise of transit candidates to be significantly lower. 

\end{itemize}

\noindent Vetters were instructed to be liberal in designating TCEs as ``New Candidates'', as part of a ``innocent until proven guilty'' approach that aimed to pass all potentially transiting planets.

A minimum of two independent human vetters were required to examine each TCE and choose a category. In the event of disagreement between the first two vetters an examination by at least one additional, independent vetter was performed. Final categories were assigned to each TCE by examining the fraction of the votes for each category. In order to be designated a ``New Candidate'', greater than 50\% of vetters had to vote for the ``New Candidate'' option. Similarly, the ``Instrumental'', ``Variable Star'', and ``Low S/N'' categories required greater than 50\% of votes to be designated as such. Of the 14,920 designated TCEs that entered Triage, 3,616 were designed as ``New Candidate'', 1,185 as ``Instrumental'', 6,566 as ``Variable Star'', 611 as ``Low S/N'', and 2,942 did not receive a majority of votes for any category.

The 3,616 TCEs designated ``New Candidate'' were subjected to an additional level of scrutiny via an independent analysis that utilized different detrending and transit modeling techniques than \citet{Tenenbaum2013} as described in \S\ref{lca}. TCEs that were found to correspond to the secondary eclipse of a system, or had too low of a signal-to-noise to be recovered by the independent analysis (typically less than $\sim$7), were not assigned KOI numbers. Only about half of all ``New Candidate'' TCEs were assigned KOI numbers. Combining the new KOIs found from the Q1-10 and Q1-12 exercises yielded a total of \nqtwelvenewkois KOIs to disposition. In Figure~\ref{tcefig} we plot these new KOIs as a function of period in blue. As can be seen in Figure \ref{tcefig}, the Triage process greatly reduced the short- and long-period TCE excesses.

It should be noted that previously, our catalogues did not assign KOI numbers to ``obvious eclipsing binaries'' \citep{Borucki2011a,Borucki2011b,Batalha2013,Burke2014}. This often included systems that showed no evidence for being an eclipsing binary other than large primary transit/eclipse depths. As stellar parameters, particularly radius, are notoriously unreliable, it raises the question as to whether large, Jupiter-sized planets around small, M-dwarf stars have been repeatedly rejected from the KOI list in past exercises. As well, it would become tedious to continually re-vet every eclipsing binary in the field for each new exercise if new KOI numbers were not assigned to them. Thus, principally for these two reasons, we assign KOI numbers to all transiting/eclipsing systems, including stellar binaries, in the Q1-Q12 exercise. However as mentioned in \S\ref{tcesec}, many known EB candidates were excluded from the pipeline run.

\subsection{Dispositioning}
\label{dispsec}

In Dispositioning, human vetters, from the Threshold Crossing Event Review Team (TCERT), were asked to determine if a KOI showed evidence for being a binary, a background eclipsing binary or instrumental artifact.  The vetters were given an electronic document with 8 pages per KOI and asked to separately evaluate the KOI according to its flux (photometric time-series) and centroid (pixel-level time-series) data. For the flux dispositioning, the vetters were asked to specify a specific reason for failure if they were sufficiently convinced the KOI was a false positive. For centroid vetting, the vetters were simply asked to choose whether or not the KOI was a planet candidate or false positive. In the following subsections we discuss each page's contents and how they were used for dispositioning.

\begin{figure*}
\centering
\includegraphics[width=\linewidth]{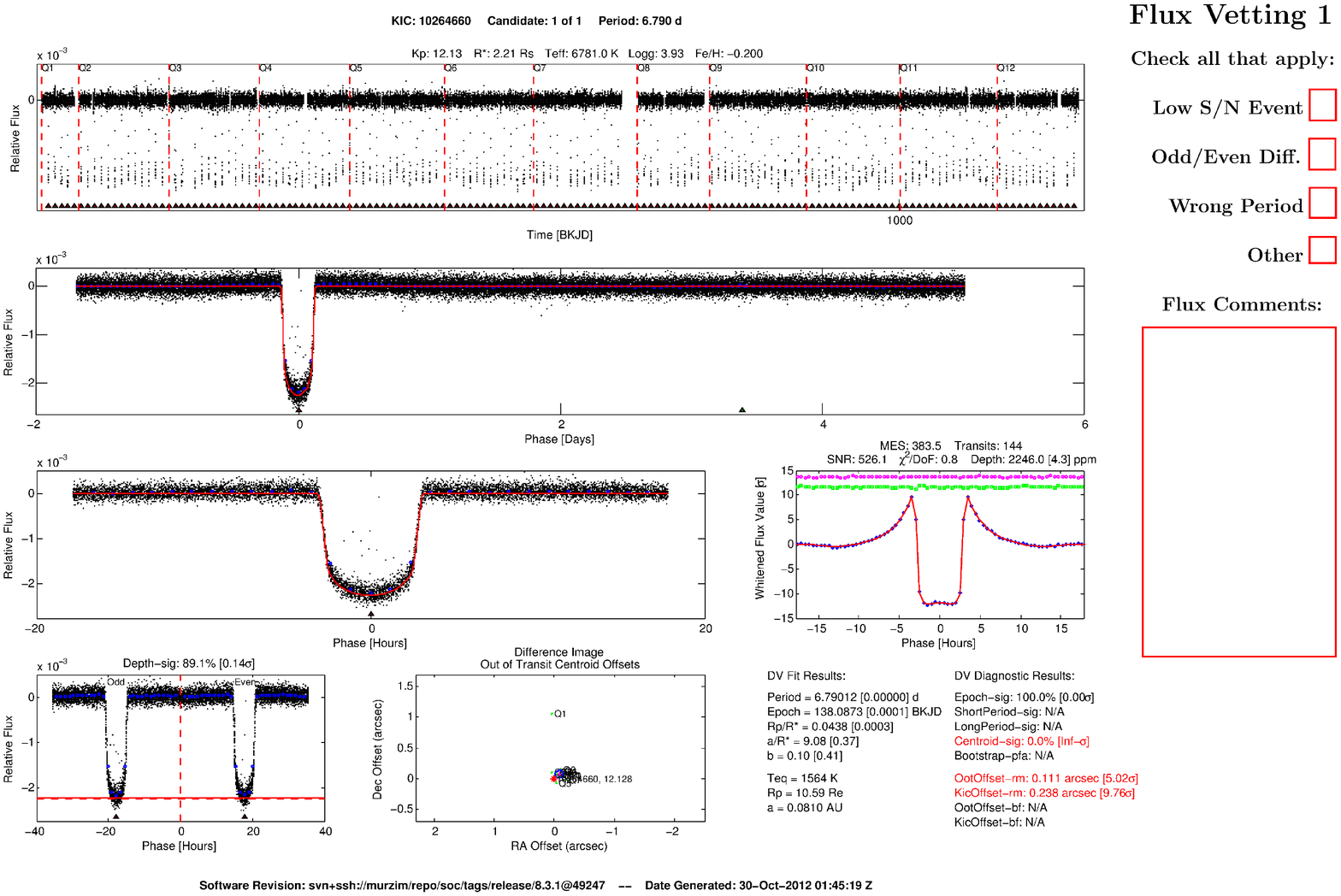}
\caption{Example of the Q1-Q12 TCERT Dispositioning form for Kepler-14b, a well-known confirmed planet. The first page, Figure~\ref{disppdffig1} is shown here in the text, with Figures~\ref{disppdffig2}-\ref{disppdffig8} showing the remaining 7 pages in the Appendix.}
\label{disppdffig1}
\end{figure*}

\subsubsection{Page 1: The Q1-Q12 DV One-Page Summary}
\label{dvsumsec}

On this page vetters were asked to choose one or more of the following FP categories if they were sufficiently convinced the KOI was a false positive:

\begin{itemize}

\item Low S/N Event: The KOI was a low signal-to-noise event. This indicates that no transit signal is readily visible by eye in the phased data.

\item Odd/Even Diff.: The KOI showed a significant difference in the depth of the odd- vs even-numbered transits. A common false-positive is an eclipsing binary system composed of two stars with nearly equal mass, size, and temperature. This type of false-positive may be detected by TPS at half the true period of the system, thus showing alternating eclipses with slightly different depths.

\item Wrong Period: The KOI appears to have been detected at the wrong period. This typically occurred at an integer ratio of the true orbital period, and principally for objects with large seasonal depth differences due to contamination.

\item Other: Any other reason that would indicate a FP not listed above. The vetters were encouraged to leave a text comment to explain the reason.

\end{itemize}

An example of the DV one-page summary is shown in Figure~\ref{disppdffig1}, and shows, on a single page:

\begin{itemize}

\item Top of Figure: The TCE/KOI's Kepler Input Catalog (KIC) number, the number of TCEs detected in the system, the period of the TCE/KOI, and the Kepler magnitude, size, temperature, surface gravity, and metallicity of the host star.

\item Top Panel: The full time-series of the DV photometric light curve. Individual quarters are denoted via dashed lines, transit locations are donated with triangles along the bottom, and the CCD module and channel number are shown in brackets alongside each quarter number.

\item Second Panel from Top: The phase-folded photometric light curve for the entire orbital cycle overlaid with binned points and the best-fit transit model.

\item Third Panel from Top on the Left: The phase-folded photometric light curve narrowed to within a couple transit durations of the primary event, also with binned points and the best-fit transit model.

\item Third Panel from Top on the Right: The whitened light curve with the best-fit whitened transit model, the residuals, and the whitened time-series at half an orbit after the transit. The numbers at the top of this panel show the detected MES, the number of transits, the S/N, reduced $\chi^{2}$, and depth of the whitened transit model fit.

\item Bottom Left Panel: The phase-folded light curve for odd- and even-numbered transits separately. The top of this panel shows a metric that indicates the similarity of the two depths (see \S 5.5 of \citealt{Rowe2014} or \citealt{Wu2010}).

\item Bottom Middle Panel: The measured centroid offset for each individual quarter and all quarters combined.  The centroid method we use is the fit of a Point Response Function (PRF) to the pixel difference image constructed by subtracting an average in-transit image from an average out-of-transit image \citep{Bryson2013}.

\item Bottom right Panel: A table of various model transit fit parameters and centroid diagnostics.  Some parameters and diagnostics can be listed as {\it N/A} when the computation was either invalid, or was not calculated.

\end{itemize}

This page was principally used to quickly assess the significance and type of the transit-like event and search for any difference in depth between the odd- and even-numbered transits. At one glance, a vetter could tell whether the TCE was due to something resembling a transiting planet, or was due to instrumental artifacts, starspots, a pulsating star, an eclipsing binary, or other phenomena. Although pixel-level centroid information and associated metrics were provided on this page, vetters were asked not to make any decisions based on them.

\subsubsection{Page 2: The Model-Shift Uniqueness Test and Occultation Search}

On this page, vetters were asked to choose one or more of the following FP categories if they were sufficiently convinced the KOI was a false positive:

\begin{itemize}

\item Transit Not Unique: The primary transit did not appear to be unique in the phased light curve. This typically occurred when there were tertiary or positive events of comparable significance to the primary event, and indicated a false alarm due to instrumental artifacts or stellar variability.

\item Secondary Eclipse: There was a significant and unique secondary eclipse event. This indicated the object was most likely an eclipsing binary with a distinct secondary eclipse.

\item Wrong Period: The KOI appears to have been detected at the wrong period. This typically occurred at an integer ratio of the true orbital period, and principally for objects with large seasonal depth differences due to contamination.

\item Other: Any other reason that would indicate a FP not listed above. The vetters were encouraged to leave a text comment to explain the reason.

\end{itemize}

We performed a uniqueness test to determine the robustness of the TCE detection and to search for secondary events.  If a KOI under investigation is truly a PC, there should not be any other transit-like events in the light curve with similar or greater depth, duration and period to the primary signal, in either the positive or negative flux directions. If such signals are present they call into question the significance of the primary event. If the primary is a unique event in the phase folded light curve, but there is also a smaller, secondary event that is unique compared to any tertiary events, then the system is most likely an eclipsing stellar binary.

Twelve quarters of data were used to search for shallow transit events (less than 100 ppm) with long periods (over 300 days).  For this type of search only a small percentage of the orbital phase contains transit information and it can be very difficult to judge the quality of a detected event when examining either a full phase-curve or a zoom-in on data close to transit. These diffculties are simply a fact of the large dynamic range of information that must be assessed to judge a transit candidate. As such, a new data product, the model-shift uniqueness test and occultation search, was developed and used in the Q1-Q12 TCERT activity to search for additional transit-like events in the data that have the same periodicity as the primary event.


To search for additional events, we took the DV photometric time series folded at the orbital period of the primary event and used the DV-generated transit model as a template to measure the amplitudes of other transit-like events at all phases. The amplitudes were measured by fitting the depth of the transit model centered on each of the data points. The deepest event aside from the primary transit event, and located at least two transit durations from the primary, was labeled as the secondary event. The next-deepest event, located at least two transit durations away from the primary and secondary events, was labeled as the tertiary event. Finally, the most positive flux event (i.e., shows a flux brightening) located at least three transit durations from the primary and secondary events was also labeled. An example is shown in Figure~\ref{disppdffig2}.

We determined the uncertainty in the amplitude measurements by calculating the standard deviation of the unbinned photometric data points outside of the primary and secondary events. Dividing the amplitudes by this standard deviation yielded significance values for the primary ($\sigma_{Pri}$), secondary ($\sigma_{Sec}$), tertiary ($\sigma_{Ter}$), and positive ($\sigma_{Pos}$) events shown at the top-left of Figure~\ref{disppdffig2}. Assuming there are $P/T_{\rm dur}$ independent statistical tests per TCE, where $P$ is the period of the KOI and $T_{\rm dur}$ is the transit duration, we computed a detection threshold for each TCE such that this test yielded no more than one false alarm when applied to all KOIs. We called this threshold $\sigma_{FA}$, and computed it via the following equation, 

\begin{equation}
\sigma_{FA} = \sqrt{2}\cdot\mathrm{erfcinv}\left(\frac{T_{\rm dur}}{P \cdot {\rm nKOIs}}\right),
\end{equation}

\noindent where erfcinv is the inverse complementary error function and nKOIs is the number of KOIs dispositioned. Finally, we also measure the amount of systematic red noise in the lightcurve on the timescale of the transit by computing the standard deviation of the measured amplitudes outside of the primary and secondary events defined by the duration of the primary event. We report the value $F_{Red}$, which is the standard deviation of the measured amplitudes divided by the standard deviation of the photometric data points. If $F_{Red}$ = 1, there is no red noise in the lightcurve. It should be noted that if no DV fit was performed for the given TCE, this plot and its associated statistics could not be generated.

The model-shift uniqueness test and occultation search was crucial in eliminating many of the false positives associated with the $\sim$372 day long-period TCE excess discussed in \S\ref{tcesec}, as well as identifying eclipsing binaries with shallow secondary eclipses.

\subsubsection{Page 3: The Centroid Vetting Summary}
\label{centsumsec1}

As {\it Kepler's} pixels are nearly 4$\arcsec$ in size and as \ik does not have an optimal point spread function across the field of view, many target KOIs are contaminated by other nearby astrophysically varying objects. In such cases, the other astrophysical signal is observed in the photometric light curve of the target KOI at a reduced amplitude. However, by examining the pixel-level data, the true source of the signal can be identified as not belonging to the target KOI, thus making the event a false positive. The remaining pages of the dispositioning document were dedicated to assisting in this determination. Here we present them and briefly discuss their use in pixel-level centroid vetting; for a comprehensive review on the identification of false positives using the pixel-level data, see \citet{Bryson2013}.

Page 3 of the DV document the centroid vetting summary page provides more in-depth pixel-level centroid information than that presented in the DV summary (see \S\ref{dvsumsec}). Three different yet complementary reconstructions of the location of the transit signal relative to the target star were presented, as shown in Figure~\ref{disppdffig3}. This page contains three elements:

\begin{enumerate}

\item Descriptive information about the target:

  \begin{enumerate}
  
  \item The Kepler magnitude, which is important in order to identify saturated targets, whose saturated pixels do not provide reliable centroiding information.  When the target star is bright enough that saturation may be an issue this value is turned red.

  \item The transit S/N as measured by the DV transit model fit. This correlates to the quality of the difference images used to measure centroid offsets displayed in the bottom-middle panel of page 1 of the DV report.

  \item The number of quarters with good difference images. This refers to the difference image quality metric, which tells how well the fitted Pixel Response Function (PRF --- \ikt's point spread function convolved with quarterly motion) is correlated with the difference image pixel data. A difference image fit was considered good if the correlation is $>$ 0.7. If the correlation is smaller this does not mean that the quarter's difference image was useless, rather that the vetter had to examine it more carefully. When the number of good quarters is three or less this line turned red.

  \item The distance from the out-of-transit PRF-fit centroid to the target star's catalog position. When this distance is $>$ 2$\arcsec$ the text was turned red, and indicated that either the catalog position or the out-of-transit PRF-fit was in error.
  
  \end{enumerate}
  
\item A table giving the reconstructed location of the transit signal relative to the target star using three different but complimentary methods:

  \begin{enumerate}
  
  \item The multi-quarter average offset of the PRF-fit difference image centroid from the PRF-fit out-of-transit (OOT) image.

  \item The multi-quarter average offset of the PRF-fit difference image centroid from the Kepler Input Catalog (KIC) position.

  \item The offset reconstructed from photometric centroids.

  \end{enumerate}

For all of these methods the distance, significance, and sky co-ordinates were reported. An offset distance was considered to be statistically significant when it was greater than 3\,$\sigma$ as well as greater than $\sim$0.1$\arcsec$. The latter condition is due to a $\sim$0.1$\arcsec$ noise floor resulting from spacecraft systematics, below which it does not appear possible to reliably measure centroid offsets.

\item Three panels showing the reconstructed location of the transit signal relative to the target star (located at 0,0), which corresponded to the three rows of the table. The first two panels, based on PRF-fitting techniques, showed the offset from the out-of-transit fit and the KIC position, respectively. In each of these panels the crosses represented each individual quarter, with the size of the crosses corresponding to their 1$\sigma$ errors. The circle was the 3$\sigma$ result for all quarters combined. The third panel showed the offset location based on photometric centroids, which provided only a multi-quarter result. The vetters were instructed to examine if any bright stars were near the target that may have influenced the PRF fit by comparing the calculated offsets from the out-of-transit PRF fit and the KIC position.

\end{enumerate}

Vetters were not asked to check boxes on this page, but to keep the information in mind for a final decision on the final page (see \S\ref{centsumsec2}).

\subsubsection{Pages 4-6: The Pixel-Level Difference Images Vetting Summary}

The next three pages showed the average difference and out-of-transit images for each quarter, which provided the data behind the PRF-fit centroids and the resulting multi-quarter average. These images were arranged so that they showed four quarters, or a full year, per page. Each image showed three positions via markers: ``x'' marked the catalog location of the target star, ``+'' marked the PRF-fit centroid of the OOT image, and ``$\Delta$'' marked the PRF-fit centroid of the difference image. The colour bar was a crucial interpretation tool: when it was almost entirely positive for the difference image, this meant that the difference image was reliable. Large negative values were marked with large, red ``X'' symbols, and indicated that the difference images were unreliable, or that the TCE was due to systematics that did not have a stellar PRF. White asterisks indicated background stars with their Kepler ID and magnitudes. This included stars from the UKIRT catalog, which had Kepler IDs $>$ 15\,000\,000. These UKIRT Kepler IDs were internal project numbers and did not correspond with UKIRT catalog identifiers. A North-East (N/E) direction indicator was provided to allow matching with the figures on page 3 (see \S\ref{centsumsec1}). Examples for Quarters 1--12 are shown in Figures~\ref{disppdffig4}, \ref{disppdffig5}, and \ref{disppdffig6}. 


Vetters were asked to denote any difference image that did not appear to be due to a stellar PRF by checking the box to the right of each quarter. If the difference image appeared to resemble a healthy looking stellar PRF, the vetters were instructed to determine if the location of the source indicated by the difference image was coincident or not with the location of the target KOI. The vetters were instructed to retain this information for a final decision on the final page of the DV document (see \S\ref{centsumsec2}).

\subsubsection{Page 7: The Flux-Weighted Photometric Centroids}

This page of the DV document showed the flux-weighted photometric centroids, which were used to confirm if the centroid shift occured at the time of transit. The top panel showed the phase-folded DV photometric time-series. The middle and bottom panels showed the computed RA and Dec centroid offsets, respectively, for each photometric data point. A photometric offset could be considered to be observed if there was a change in the centroid time series (second and third panel) that looked like the flux time series (top panel). The purpose of this figure was to verify that if there was a measured photometric shift from the difference images, it looked like the transit signal, and thus was not due to instrumental systematics or stelalr varibility.  Vetters were asked to mark a box at the top of the page if there was significant signal in the photometric centroids, but it did not resemble the transit shape. An example is shown in Figure~\ref{disppdffig7}. It should be noted that vetters were instructed to never fail a KOI based on the photometric centroids alone as a photometric centroid shift in transit does not itself imply an offset source and the chances of being a false positive are much higher when the centroids are unresolved, particularly at low Galactic latitudes \citep{Bryson2013}.

\subsubsection{Page 8: The Centroid Vetting Summary With Checkboxes}
\label{centsumsec2}

The last page of the form, Page 8, was a repeat of page 3, but with final decision checkboxes added on, as shown in Figure~\ref{disppdffig8}. Here the vetters were asked to select one of the following options:

\begin{itemize}

\item Pass: The pixel-level data indicated that the source of the transit-like signal was coincident with the target KOI, and thus the KOI was a planet candidate.

\item Maybe: The pixel-level data was not conclusive, and the vetter did not feel comfortable making a decision.

\item No Data: There was not sufficient information to determine the location of the source of the transit-like signal, either due to a lack of a fitted transit model or very low signal-to-noise. This option designates the KOI as a planet candidate, but is recorded separately from ``Pass'' for data analysis purposes.

\item Fail: The location of the transit signal does not coincide with the location of the target KOI, thus the KOI is a false positive.

\end{itemize}

For flux vetting, if any false positive reason was marked by a vetter the KOI was considered a flux fail by that vetter, else it was considered a flux pass. For centroid vetting, if ``Fail'' was marked by a vetter the KOI was considered a centroid fail, if ``Pass'' or ``No Data'' were marked the KOI was considered a centroid pass, and if ``Maybe'' was marked the KOI was considered not to have been centroid vetted. Similar to Triage (see \S\ref{triagesec}), a minimum of two independent human vetters were required to examine each KOI and vet both flux and centroids. If the two vetters disagreed on a pass or fail disposition for the flux and/or centroid portions, examination by at least one additional, independent vetter was performed. Final pass/fail categories were assigned to each KOI for their flux and centroid data. In order to be designated a ``Planet Candidate'' the KOI had to pass both the flux and centroid vetting. If the KOI failed either portion, or both, it was designated a ``False positive''.  The reasons for dispositions assigned through Q1-Q12 activities are available at the NASA exoplanet archive.

\subsection{Ephemeris Matching}
\label{ephemmatchsec}

In a parallel activity to the TCERT vetting an effort was made to examine the periods and epochs of all known KOIs and eclipsing binaries within the \ik field of view known from both space- and ground-based observations. In short, if a KOI is contaminated from another source, their ephemerides (periods and epochs) will be nearly identical. Thus, false positive KOIs may be identified by simply matching their periods and epochs to other KOIs and EBs. A thorough matching of these ephemerides for all KOIs in the Q1-Q12 catalog, along with previous catalogs, was performed and the work fully documented in a separate paper \citep{Coughlin2014}. As a result, 685 KOIs were identified as false positives, some of which were among the KOIs vetted by the Q1-Q12 TCERT activity, and some of which were dispositioned in previous catalogs.

Over 100 of these false positive KOIs were not identified as such by either the Q1-Q12 TCERT activity or previous vetting activities. These are predominately low signal-to-noise KOIs that have been contaminated by sources many tens of arcseconds away, such that no clear centroid offset is observed, as the KOI lies in the far wings of the contaminating source's PRF. The ability to identify these cases and study them will lead to improved metrics and procedures for identifying these cases in the future.

\section{Planet Candidate Sample}\label{pccandidate}

As a result of the TCERT vetting, including triage, dispositioning, and ephemeris matching, we dispositioned \nvettedkois KOIs as \nqtwelvevettedpcs PCs and \nqtwelvevettedfps FPs.  These KOIs and their new dispositions are available at the NASA Exoplanet Archive\footnote{http://exoplanetarchive.ipac.caltech.edu/}. We augment this definition of a PC by also requiring that the modeled signal-to-noise ratio of the detected transit with Q1-Q17 ($\sim$\,4 years) photometry be greater than 7.1.  We further retain all KOIs that have dispositions labeled as {\it CONFIRMED} in the NASA Exoplanet Archive, except KOI-245.04 which is a known false-alarm \citep{Barclay2013}.  This brings the total number of designated KOIs to \nallkois and the total number of designated PCs to \nallpcsk. 

In the NExScI archive, we also include 4 flag columns to indicate the reasons a KOI was marked as a false positive. The flags indicate if a KOI was determined to be:

\begin{itemize}
\item ``Not Transit-like'': A KOI whose light curve is not consistent with that of a transiting planet. This includes, but is not limited to, instrumental artifacts, non-eclipsing variable stars (e.g., heartbeat stars, \citet{Thompson2012}), and spurious detections.

\item ``Significant Secondary'': A KOI that is observed to have a signicant secondary event, meaning that the transit event is most likely caused by an eclipsing binary.

\item ``Centroid Offset'': The source of the transit was on a nearby star, not the target KOI.

\item ``Ephemeris Match Indicates Contamination'': The KOI shares the same period and epoch as another system and is judged to be a false positive as described in \S\ref{ephemmatchsec}.
\end{itemize}

\noindent More than one flag can be set simultaneously, and no flags are exclusive, although generally a KOI was never failed as both due to ``Not Transit-Like'' and ``Significant Secondary''. The only cases in which both of those flags are set are cases where a KOI number was accidentally designated to correspond to the secondary eclipse of a system.

In Figure~\ref{tcefig} we plot a period histogram that includes the Q1-Q12 TCE population in red, the \nallkois KOIs known in green, the \nqtwelvenewkois new KOIs designated by Q1-Q12 TCERT in blue, and the \nqtwelvenewpcs new PCs as a result of the Q1-Q12 vetting activity in cyan. As can be seen, the final population of planet candidates do not exhibit any short- or long-period excess due to false positives, thus validating the effectiveness of our tests. Compared to previous catalogs, while we have added new PCs at all periods, we have especially augmented the sample of PCs at long periods.

\subsection{Stellar Parameters}
\label{stellarpars}

\begin{figure}[h]
\centering
\begin{tabular}{c}
\includegraphics[width=\linewidth]{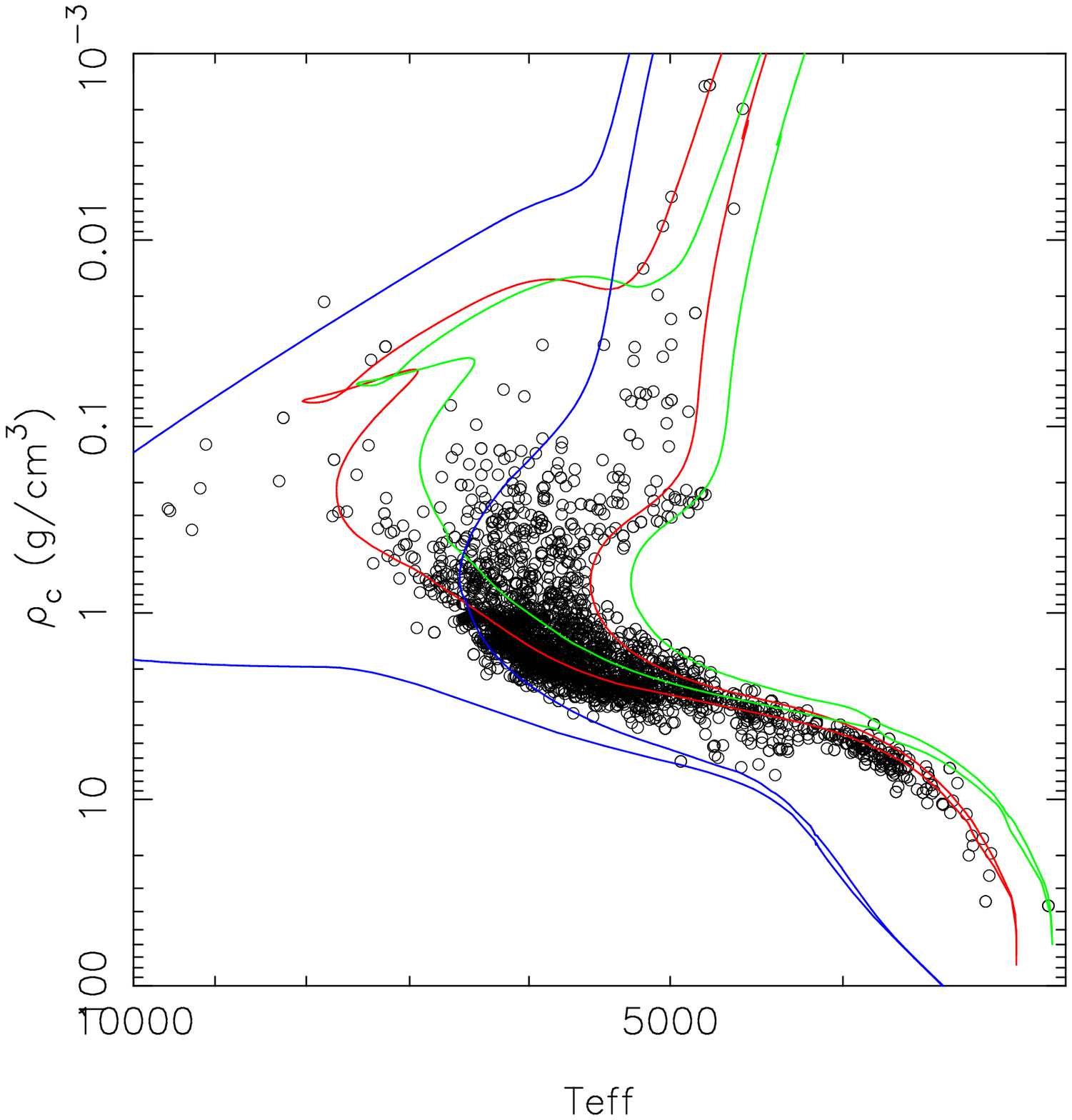} \\
\includegraphics[width=\linewidth]{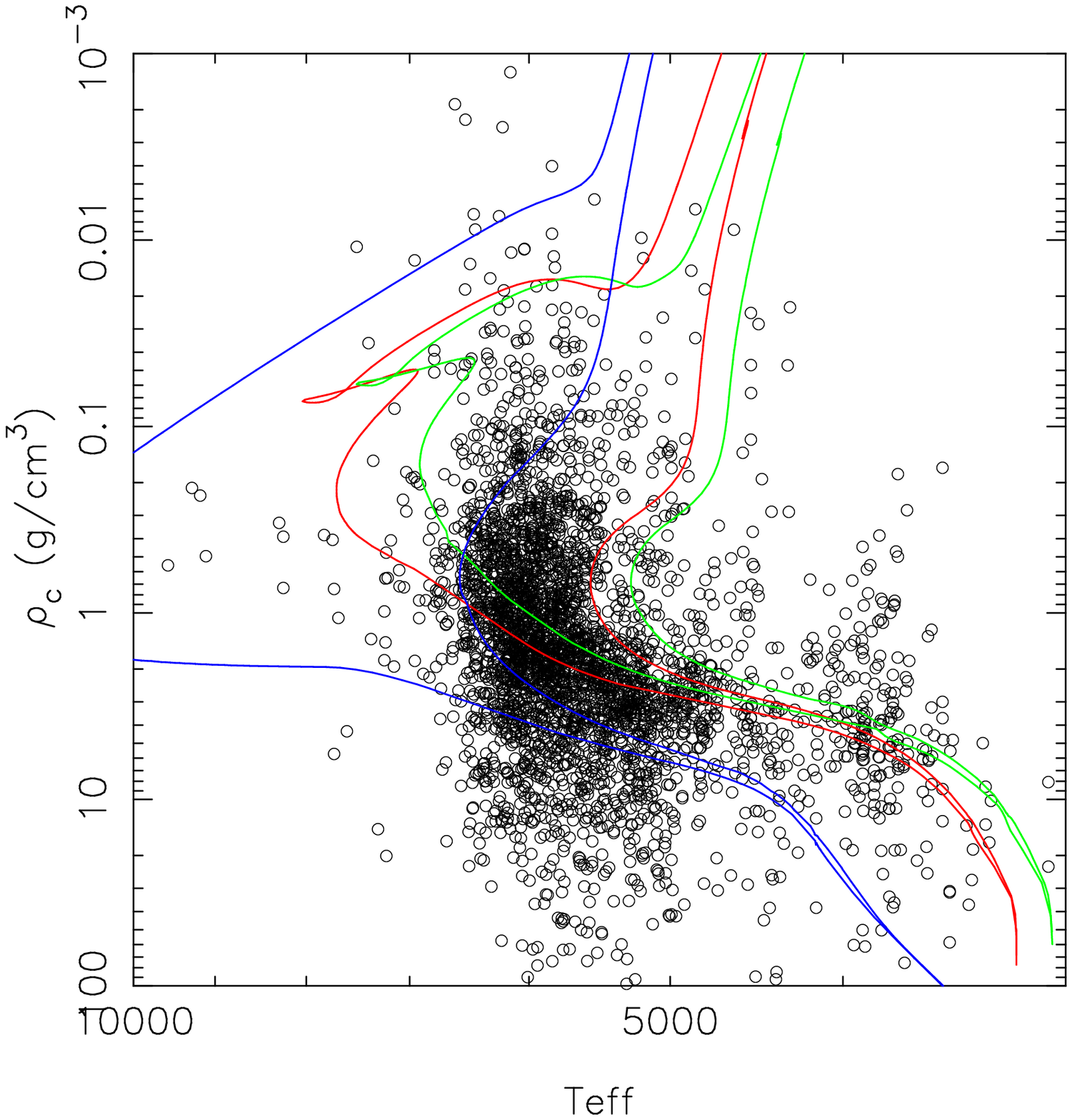} 
\end{tabular}
\caption{The top panel shows the adopted stellar parameters plotted as mean stellar density (\rhostar) vs \teff\  from \cite{Huber2014} for planetary-candidates.  The bottom panel shows the inferred mean stellar density based on our adopted circular orbit transit models (\rhoc) for planetary candidates. The red lines show Dartmouth isochrones with Solar metallicity with ages of 1 and 14 Gyr, the blue lines for \feh = --2.0 and the green lines for \feh = +0.5}
\label{hrdiag}
\end{figure}

Our adopted stellar parameters are based on \citet{Huber2014}, which uses atmospheric parameters (\teff, \logg, \feh) derived from a variety of observation techniques such as photometry, spectroscopy and asteroseismology that are homogeneously fit to the grid of Dartmouth stellar isochrones \citep{Dotter2008} to estimate the stellar mass and radius (\mstar\ and \rstar).   The top panel of Figure \ref{hrdiag} displays our adopted stellar parameters for PCs as defined in \S \ref{pccandidate}.   Overlaid are Dartmouth isochrones with ages of 1 and 14 Gyrs and \feh = -2.0 (blue), 0.0 (red) and +0.5 (green).  {\it Kepler's} PCs are found preferentially around dwarf stars as opposed to evolved giants.  This is expected as transit depth is directly proportional to the ratio of the planet and star radius (\rprs).  There is also a noted lack of PCs with host stars hotter than $\sim$6500 K, which is due to \ik mission target selection \citep{Batalha2010}, increasing stellar radius with \teff\ across the Zero-Age-Main-Sequence (ZAMS) and pulsational properties of A and F-stars.  

Our stellar parameters for both PC and FP target stars are listed in Table \ref{spars}.  The stellar characterizations are used to derive our measured fundamental parameters of PCs as described in \S \ref{lca}.

\begin{deluxetable*}{cccccccccc}
\tabletypesize{\scriptsize}
\tablecaption{Stellar Parameters}
\tablewidth{0pt}
\tablehead{
\colhead{KOI} & \colhead{\teff} & \colhead{\teff$\sigma$} & \colhead{\logg} & \colhead{\logg$\sigma$} & \colhead{\feh} & \colhead{\feh$\sigma$} & \colhead{\rstar} & \colhead{\rstar+$\sigma$} & \colhead{\rstar-$\sigma$} \\
\colhead{} & \colhead{K} & \colhead{K} & \colhead{cgs} & \colhead{cgs} & \colhead{} & \colhead{} & \colhead{\rsun} & \colhead{\rsun} & \colhead{\rsun}
}
\startdata
1 & 5850 & 50 & 4.455 & 0.025 & -0.150 & 0.100 & 0.950 & 0.020 & -0.020 \\
2 & 6350 & 80 & 4.021 & 0.011 & 0.260 & 0.080 & 1.991 & 0.018 & -0.018 \\
3 & 4777 & 92 & 4.590 & 0.026 & 0.320 & 0.120 & 0.765 & 0.030 & -0.022 \\
4 & 6244 & 120 & 3.657 & 0.156 & -0.160 & 0.170 & 2.992 & 0.469 & -0.743 \\
5 & 5753 & 75 & 4.003 & 0.011 & 0.050 & 0.101 & 1.747 & 0.042 & -0.042 \\
6 & 6178 & 118 & 4.106 & 0.164 & 0.000 & 0.130 & 1.580 & 0.415 & -0.340 \\
7 & 5781 & 76 & 4.105 & 0.010 & 0.090 & 0.101 & 1.533 & 0.040 & -0.040 \\
8 & 5842 & 115 & 4.433 & 0.109 & -0.100 & 0.150 & 0.985 & 0.187 & -0.079 \\
9 & 6277 & 169 & 4.457 & 0.176 & -0.220 & 0.270 & 1.015 & 0.465 & -0.090 \\
10 & 6225 & 136 & 4.169 & 0.051 & -0.040 & 0.140 & 1.451 & 0.117 & -0.129 \\
11 & 5514 & 147 & 4.569 & 0.141 & -0.040 & 0.260 & 0.826 & 0.306 & -0.059 \\
12 & 6635 & 178 & 4.193 & 0.235 & -0.040 & 0.305 & 1.516 & 0.885 & -0.330 \\
13 & 9107 & 341 & 3.867 & 0.192 & 0.070 & 0.395 & 3.031 & 1.198 & -0.944 \\
14 & 8090 & 278 & 4.002 & 0.220 & -0.140 & 0.320 & 2.234 & 1.034 & -0.495 \\
15 & 8605 & 316 & 4.195 & 0.246 & 0.210 & 0.360 & 1.878 & 1.340 & -0.268
\enddata
\tablecomments{Table~\ref{spars} is published in its entirety in the electronic edition of the Astrophysical Journal. A portion is shown here for guidance regarding its form and content.}
\label{spars}
\end{deluxetable*}

\section{Transit Models}
\label{lca}

We modeled the observed transits with Q1-Q17 long-cadence photometry downloaded from the MAST\footnote{Observations labeled as PDC\_FLUX from FITS files retrieved from the Mikulski Archive for Space Telescopes (MAST) based on Data Releases 21-23.}  archive.  The photometry includes systematic corrections for instrumental trends and estimates of dilution due to other stars that may contaminate the photometric aperture \citep{Stumpe2014}.  The median value of light contamination for validated \ik planets is $\sim$5\% \citep{Rowe2014}.  We do not attempt to compensate for stellar binarity, thus in cases such as KOI-1422 (Kepler-296) our reported planetary radius is underestimated \citep{Lissauer2014,Star2014}.

We adopted the photometric model described in \S4 of \citet{Rowe2014} which uses a quadratic limb-darkened model described by the analytic model of \citet{Mandel2002} and non-interacting Keplerian orbits.  We account for gravitational interactions of planetary orbits by measuring transit-timing variations (TTVs) and including the effects in our transit models as described in \S4.2 of \citet{Rowe2014}.  Measured TTVs for all KOIs are listed in Table \ref{ttvcat}.  The model was parameterized by the mean-stellar density (\rhostar), photometric zero point and for each planet ($n$) an epoch (T$0_n$), period ($P_n$), scaled planetary radius (\rprs$_n$) and impact parameter ($b_n$).  The scaled semi-major axis for each planet candidate is estimated by
\begin{equation}\label{rhostar}
\left( \frac{a}{\rstar} \right)^3 \simeq \frac{\rhostar G P^2}{3 \pi}.
\end{equation} 
It is important to note that Equation \ref{rhostar} assumes that the sum of the planetary masses is much less than the mass of the host star.  For a 0.1 \msun\ companion of a Sun-like star, a systematic error of 2\% is incurred on the determination of \rhostar

\begin{deluxetable}{cccc}
\tabletypesize{\scriptsize}
\tablecaption{TTV Measurements}
\tablewidth{0pt}
\tablehead{
\colhead{$n$}  & \colhead{$t_n$} & \colhead{$TTV_n$}  & \colhead{$TTV_{n\sigma}$} \\
\colhead{}     & \colhead{}      & \colhead{days}     & \colhead{days} 
}
\startdata
KOI-1.01 & & & \\
1 & 55.7633008 & 0.0000101 & 0.0000626 \\
2 & 58.2339142 & 0.0001149 & 0.0000945 \\
3 & 60.7045276 & 0.0000519 & 0.0000662 \\
4 & 63.1751410 & 0.0000277 & 0.0000740 \\
5 & 65.6457543 & -0.0001186 & 0.0000623 \\
6 & 68.1163677 & -0.0000427 & 0.0000423 \\
7 & 73.0575945 & -0.0000419 & 0.0000747 \\
8 & 75.5282079 & 0.0001032 & 0.0000745 \\
9 & 77.9988213 & 0.0000349 & 0.0000673 \\
10 & 80.4694347 & -0.0000052 & 0.0000506 \\
... & & &
\enddata
\tablecomments{Table \ref{ttvcat} is published in its entirety in the electronic edition of the {\it Astrophysical Journal Letters}.  A portion is shown here for guidance regarding its form and content.}
\label{ttvcat}
\end{deluxetable}

To model the light curve, we applied a polynomial filter to the PDC flux corrected aperture photometry as described in \S4 of \citet{Rowe2014}.  This filter strongly affects all signals with timescales less than 2 days and is destructive to the shape of a planetary transit, thus we masked out all observations taken within 1 transit-duration of the measured center of the transit time and used an extrapolation of the polynomial filter.  A best fit model was calculated by a Levenberg-Marquardt chi-square minimization routine \citep{More1980} and included TTVs when necessary.  In the case of light curves that display multiple transiting candidates, we produce a light-curve for each individual candidate where the transits of the other planets were removed using our multi-planet model.  We then fit each planet individually with this light curve and use the resulting calculation to seed our Markov Chain Monte Carlo (MCMC) routines to measure fundamental physical properties of each planet.


\subsection{Model Parameters and Posterior Distributions}\label{mcmc}

Our measured planetary parameters are listed in Table \ref{mplanetfit} and are based on our transit model fits and MCMC analysis.  For multi-planet systems, each transiting planet is fitted independently . We assumed a circular orbit and fit for T0, $P$, $b$, \rprs\ and \rhoc, where \rhoc\ is the value of \rhostar\ when a circular orbit is assumed.  Thus, each planet candidate provides an independent measurement of \rhoc.  If the value of \rhoc\ is statistically the same for each planet candidate, then the planetary system is consistent with each planet being in a circular orbit around the same host star. 

To estimate the posterior distribution on each fitted parameter, we use a MCMC approach similar to the procedure outlined in \citet{Ford2005} and implemented in \citet{Rowe2014}.   Our algorithm uses a Gibbs sampler to shuffle the value of parameters for each step of the MCMC procedure with a control set of parameters to approximate the scale and orientation for the jumping distribution of correlated parameters as outlined in \citet{Gregory2011}.   Our method allows the MCMC approach to efficiently sample parameter space even with highly correlated model parameters.   We generated Markov Chains with lengths of 100\,000  for each PC. The first 20\% of each chain was discarded as burn-in and the remaining sets were combined and used to calculate the median, standard deviation and $1\sigma$ bounds of the distribution centered on the median of each modeled parameter.  Our model fits and uncertainties are reported in Table \ref{mplanetfit}.  We use the Markov Chains to derive model dependent measurements of the transit depth (T$_{dep}$) and transit duration (T$_{dur}$). The transit depth posterior was estimated by calculating the transit model at the center of transit time ($T0$) for each set of parameters in the Markov Chain.  We also convolve the transit model parameters with the stellar parameters (see \S\ref{stellarpars}) to compute the planetary radius, \rpl, and the flux received by the planet relative to the Earth ($S$).  To compute the transit duration, we used Equation 3 from \citet{Seager2003} for a circular orbit,
\begin{equation}\label{eq:tdur}
{\rm T}_{dur} = \frac{P}{\pi} \arcsin \left(  \frac{\rstar}{a} \left[ \frac{(1+\frac{\rpl}{\rstar})^2-(\frac{a}{\rstar}\cos i)^2}{1-\cos^2i} \right]^{1/2}  \right),
\end{equation}
which defines the transit duration as the time from first to last contact.  We estimate the ratio of incident flux received by the planet relative to the Earth's incident flux,
\begin{equation}
S = \left( \frac{\rstar}{\rsun} \right)^2 \left( \frac{\teff}{T_{{\rm eff}\sun}} \right)^4 \left( \frac{a}{a_{\earth}} \right)^{-2}
\end{equation}
\noindent where \teff\ is the effective temperature of host star, $T_{{\rm eff}\sun}$ is the temperature of the Sun, $a_{\earth}$ is the Earth-Sun separation and $a$ is the semi-major axis of the planet calculated with Kepler's Third Law using the measured orbital period and estimated stellar mass.  

We attempted a MCMC analysis on all KOIs, but, there are scenarios when our algorithm failed, such as when the S/N of the transit was very low (typically below $\sim$7).  In these cases, such as KOI-5.02 which is a false alarm (FA), we only report best-fit models in Table \ref{mplanetfit}.  There are no PCs without reported uncertainties.  Figure \ref{HZplot} shows an example of two parameters, $S$ and \rpl\ with uncertainties derived from our MCMC analysis.  It is common for parameters to have high asymmetric error bars.

\section{Discussion}

Based on TCERT dispositions and updates from confirmed \ik planets in the literature we list in Table \ref{mplanetfit} all \nallpcs PCs known after the Q1-Q12 vetting.  However, there are a significant number (few hundred) of PCs that have a high probability of being FPs.  The most common type of FP is an eclipsing binary in an eccentric orbit where only the primary or secondary event is seen.  The transits for these events are typically deep ($>$ 2\%) and ``V" shaped.  Our transit models suggest that many of these PC have radii larger than twice Jupiter.  However, TCERT does not disposition KOIs as FPs based on planetary radii.  Inferred radii of transiting planets depend on the stellar radius which for an individual star may incur unaccounted for large systematic error.  The DV transit model does not handle impact parameters greater than 1, which also produces systematic errors in the measured value of \rprs.  It is also unclear what the maximum radius of a planet can be due to unknown internal composition and structure and influences of external energy sources.  With our transit models and realistic posteriors, which can handle high-impact parameter cases, we now examine the PC and FP population and suggest appropriate cuts for generating a list of PCs that better represent the true exoplanet population.  At a minimum, we recommend cuts based on S/N and \rpl\ with the understanding that a few bona fide extrasolar planets will be excluded.

\subsection{Signal-to-Noise}\label{snr}

We estimate the S/N of the observed transit by estimating the noise in the photometric light curve from the standard deviation ($\sigma$) of the detected light-curve with out of transit observations compared to the transit model,
\begin{equation}
{\rm S/N} = \sqrt{\sum_{i=1}^{n} \left(\frac{Tm_i - 1}{\sigma} \right)^2},
\end{equation}
where $Tm_i$ is the value of the transit model for each observation, $i$.  The careful analysis of \citet{Fressin2013} shows that below a S/N $\sim$ 10, the detection of KOIs becomes unreliable.  Since our KOIs were depositioned by human eyes there is a tendency to keep a low S/N event that may simply be red-noise in the light-curve.  We consider all KOIs with a S/N less than 7.1 to be considered false-alarms and caution users of the KOI catalogue that all PCs with a S/N less than 10 have a significant probability of being a false-alarm.    An example is KOI-4878.01, a potentially exciting Earth-sized planet with a 450d period and a S/N of 8.  Our MCMC analysis did not consider these events to be unique, with chains jumping to other local minima.  Thus, we concluded that this KOI is likely to be a false-alarm.  If uncertainties in Table \ref{mplanetfit} are not reported, then the KOI is either a low S/N false-alarm or a transit-like event from other astrophysical processes.  {\it We recommend marking all KOIs with a S/N $<$ 7.1 or missing posteriors in Table \ref{mplanetfit} as FAs and to treat all KOIs with a S/N $<$ 10 with caution.}

\subsection{Dissecting the KOI Population}\label{dissect} 

\begin{figure}[h]
\centering
\begin{tabular}{c}
\includegraphics[width=\linewidth]{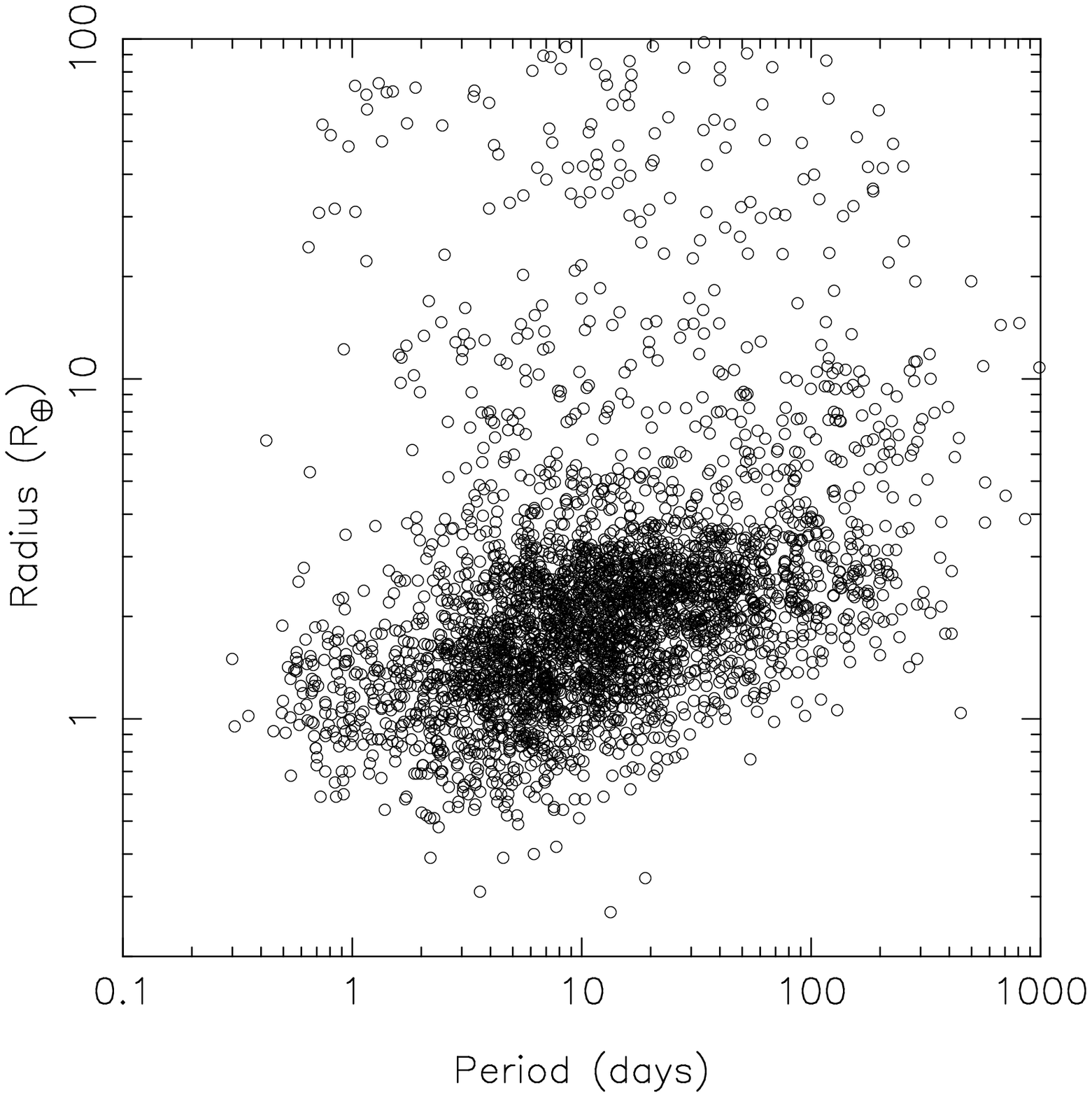} \\
\includegraphics[width=\linewidth]{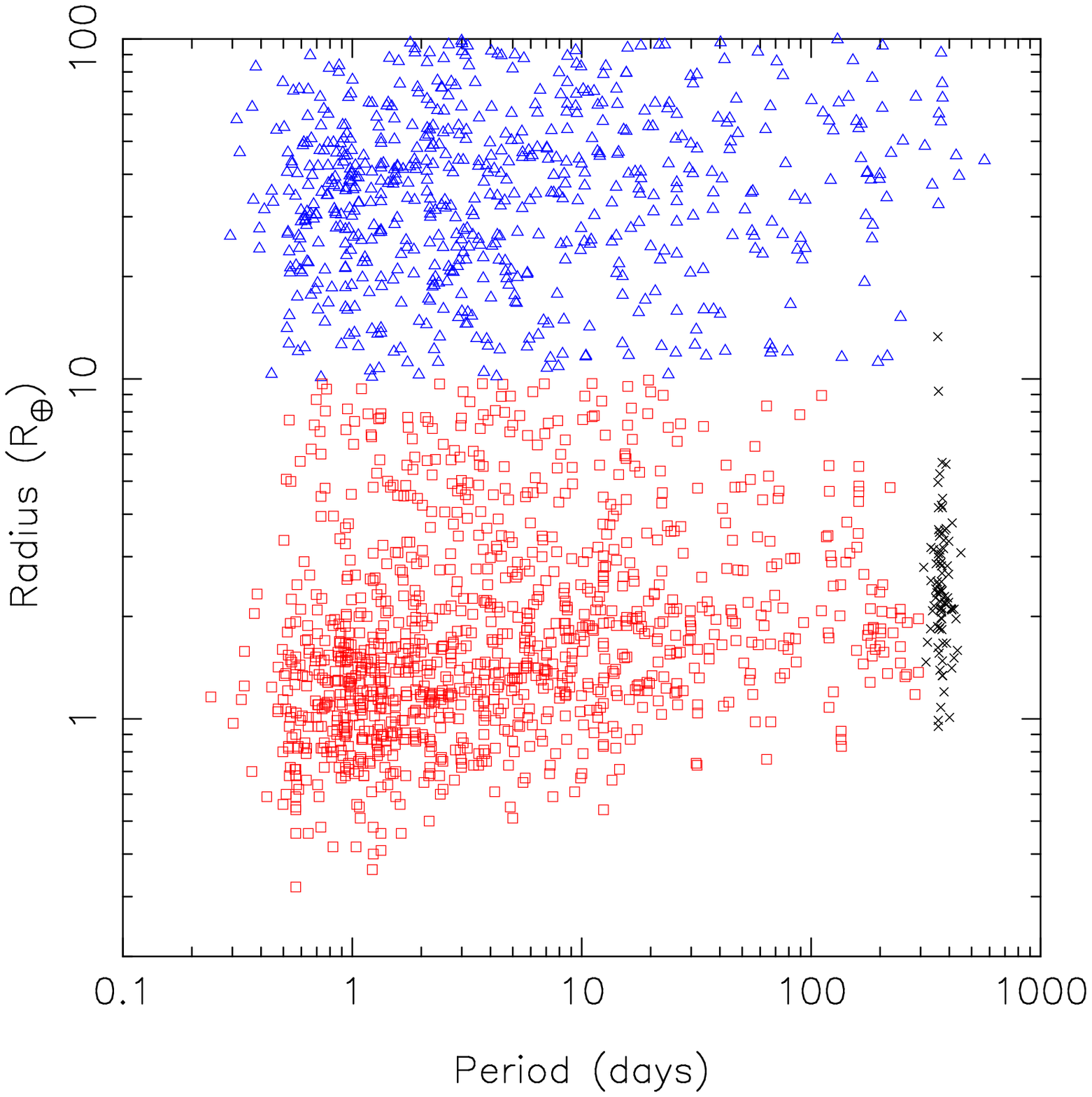} 
\end{tabular}
\caption{Period vs Radius diagram for \ik KOIs.  The top panel shows PCs and the bottom shows FPs as dispositioned by TCERT.  PCs with a radius larger than 20 \rearth\ are likely dominated by EBs.  The FP population has been divided to show objects larger and smaller than 10 \rearth\ with blue triangles and red squares respectively.  The population of FPs at 372 days is due to rolling-band instrumental systematic noise and are marked with black 'x's.}
\label{PerRad}
\end{figure}

The top panel of Figure \ref{PerRad} shows the PC population and the bottom panel shows the FP population based on TCERT dispositions.  A substantial majority of the PCs have a radius smaller than 10 \rearth, but there are 196 PCs with radius larger than even 20 \rearth\ which is larger than planetary evolution models of non-radiated, core-less, Jupiter-massed planets with ages greater than $\sim$\,100 Myr \citep[e.g.,][]{Baraffe2003}.  To produce radii above 20 \rearth\ an additional energy source is required, such as hydrogen burning present in the cores of main-sequence stars.  Thus, a majority of the PCs with radii greater than 20 \rearth\ are members of the intrinsic EB population observed by \ikt, but there will be cases where the large inferred radius is due to incorrect stellar parameters.

The bottom panel shows that the FP population can be separated into an intrinsic EB and blended EB (background eclisping binary; BGEB)\footnote{We consider BGEBs and background transiting planets to be both FPs.} population roughly divided by KOIs with transit modeled radii greater or less than $\sim$\,10\rearth.  There is also a large population of FPs centered around orbital periods of 372 days due to rolling-band instrumental systematic noise as described in \S2.1.  A BGEB is an eclipsing binary in the photometric aperture where the light is dominated by a brighter unassociated star where the two objects just happen to be aligned along the same line of sight.  The strong dilution creates a stellar eclipse that is observed to be too shallow and our transit model infers a radius that is, likewise, too small.  Even in the case of an EB, the light from the eclipsing star can be sufficient to dilute the depth of the eclipse such that the inferred radius from our planet transit model is underestimated.

The number of both EB and BGEBs decreases with orbital period due to the decreasing eclipse probability.   This is also seen for the PC population for radii smaller than $\sim$10 \rearth\ and periods greater than 2 days.  Below 2 days the planet population is likely affected by processes of planet formation and planet evaporation \citep[e.g.,][]{Owen2013}.  The change in the relative number of PCs vs BGEB for periods less than $\sim$2 days was noted in \citet{Lissauer2014} and we reiterate that point here.  Short orbital period and short transit durations combined with the \ik 30 minute observation cadence make it difficult to distinguish an EB or variable star from a transiting planet using just the \ik light curve.  The chances of a closely aligned blend that could not be detected through centroid offsets is also greatly increased due to the increasing number of EBs seen at short orbital periods.  There are projects that are successfully identifying bonafide exoplanets in this regime \citep{Sanchis2014}.

The portion of PC population with radii larger that $\sim$10 \rearth\ shows an increase in the number of candidates for periods greater than 10 days.    This is due to eccentric orbits where a secondary or primary eclipse of an EB is not seen and becomes increasingly common for longer periods and larger orbital separations.  It is possible that the stellar classification of the host star is in error for a few of these candidates.   {\it We strongly recommend that anyone using the \ik PCs apply a radius cut to eliminate the largest TCERT classified PCs.}  As an example, one could exclude all PCs above 20 \rearth\ to maintain the hot-jupiter population and accept a $\sim$35\% FP rate for Jupiter-sized planets at all orbital periods \citep{Santerne2012} due to difficulties distinguishing between late M-dwarfs, brown-dwarfs and Jupiter-sized planets.

\begin{figure}[h]
\centering
\begin{tabular}{c}
\includegraphics[width=\linewidth]{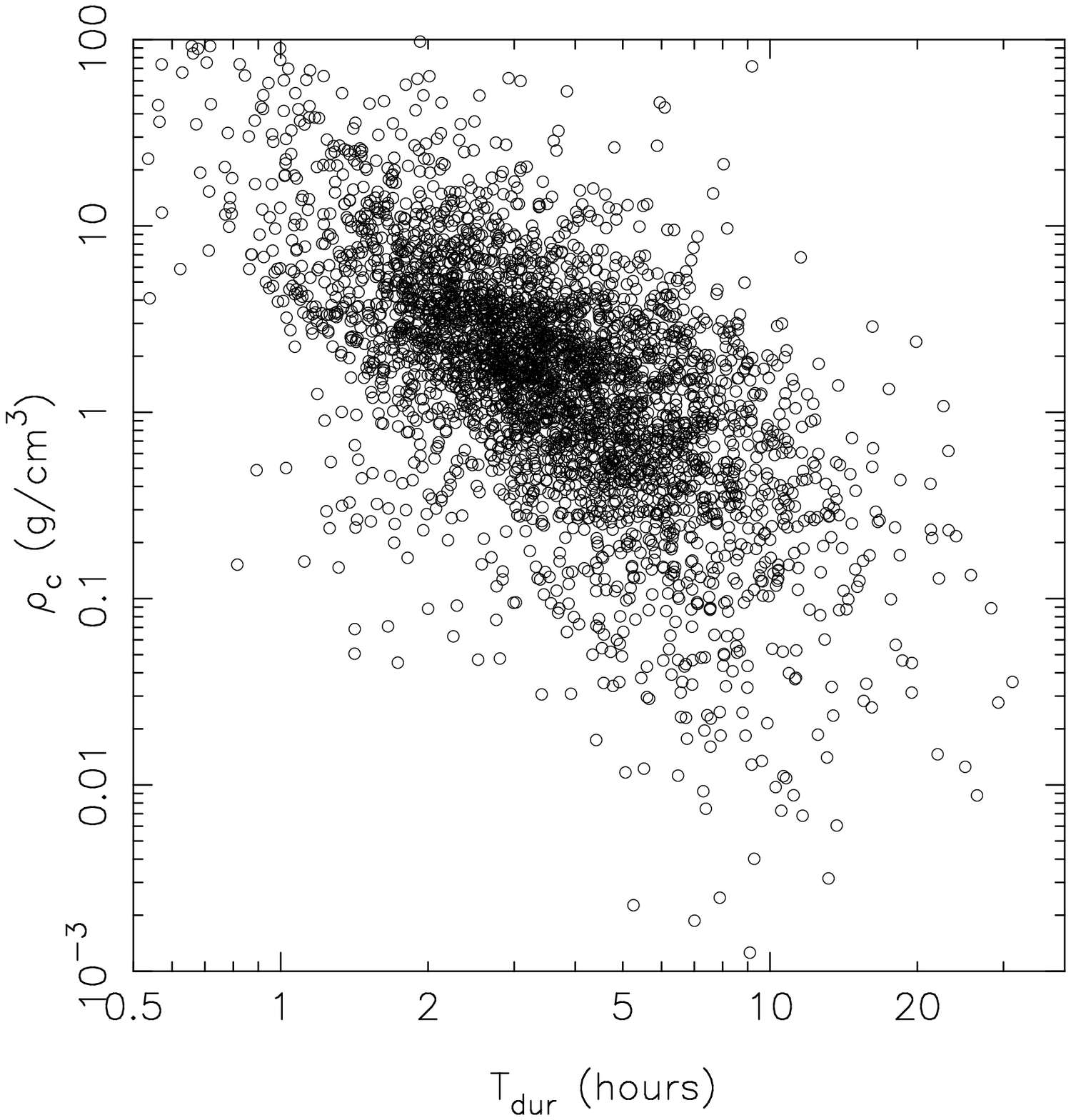} \\
\includegraphics[width=\linewidth]{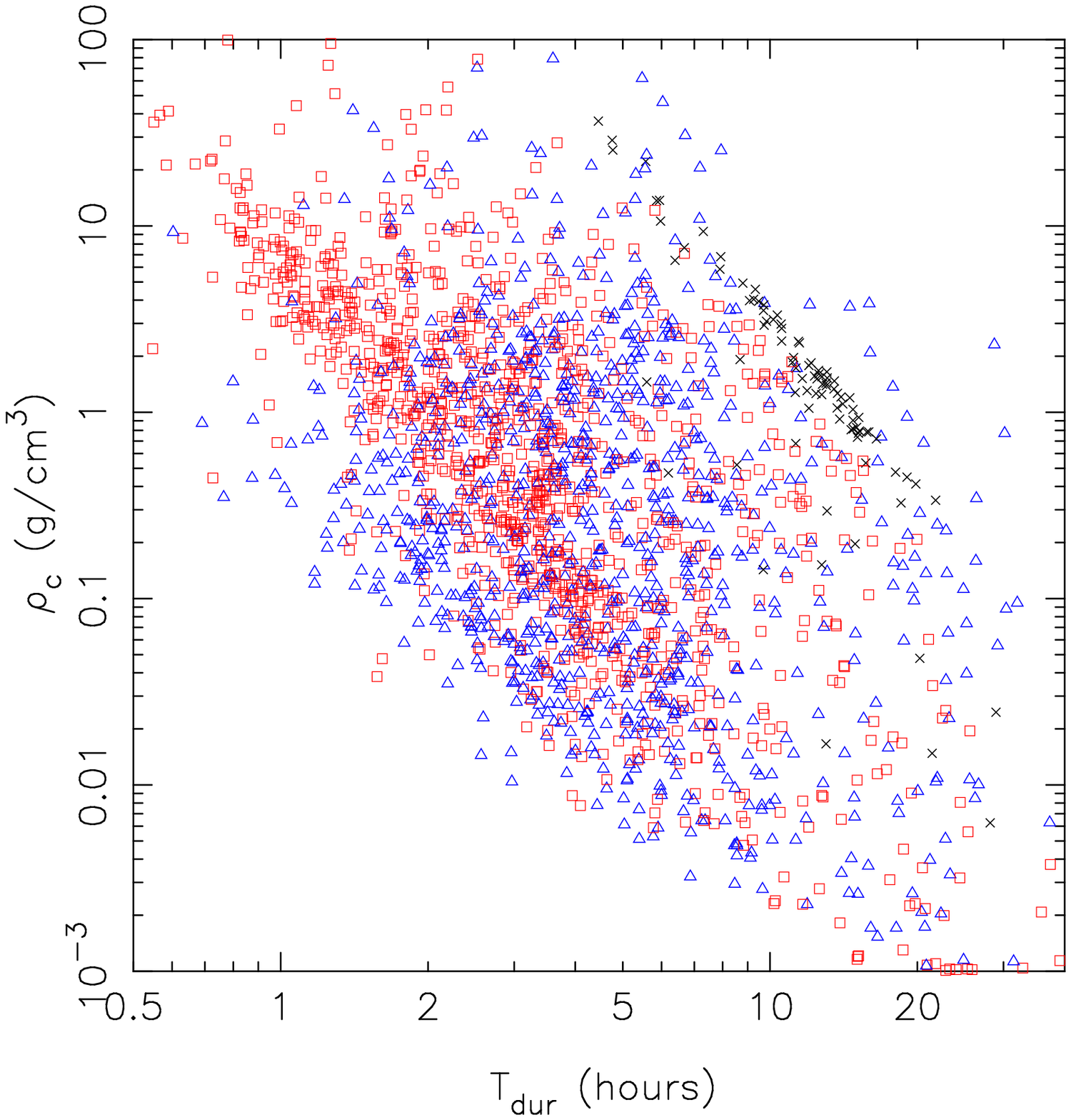} 
\end{tabular}
\caption{Transit duration vs \rhoc\ as derived from transit models.  The top panel shows PCs and the bottom shows FPs as dispositioned by TCERT.  The PC population shows an expected relation between duration and stellar properties.  As in Figure \ref{PerRad}, the FP population has been divided to show objects larger and smaller than 10 \rearth\ with blue triangles and red squares respectively to represent the EB and BGEB population.  The population of FPs near 372 days is due to rolling-band instrumental systematic noise and are marked with black `x's.  See further discussion of the FP populations in \S \ref{dissect}.}
\label{tdurrhostar}
\end{figure}

Figure \ref{tdurrhostar} displays the determined transit-duration (based on Equation \ref{eq:tdur}) and mean-stellar density for a circular orbit (\rhoc) for PCs (top panel) and FPs (bottom panel).  For planets in circular orbits around main-sequence stars it is expected that transits with shorter durations will be found around smaller, cooler stars and this correlation can be seen for the PCs.   The spread in the correlation will be due to measurement error, orbital period, impact parameter, (where a grazing transit will be shorter in duration compared to a central transit), and eccentricity (where orbital speed will vary through out the orbit).   

For the FPs, there are three populations visible.  The first can be see as a line of objects marked with `x's centered at a duration of 15 hours and a mean stellar density (\rhoc) of 2 \gcmc.  These are the FPs associated with the rolling band instrumental noise.  These candidates have a similar amplitude and period which produce a pattern that can be reproduced by Equation \ref{eq:tdur}.   

The second FP population can be seen as a cloud of FPs that extends from $T_{dur}=1$ hr, \rhoc\ = 10 \gcmc\ to $T_{dur}=10$ hr, \rhoc\ = 0.01 \gcmc\ marked with red squares.  This is the BGEB population.  It is offset towards smaller values of \rhoc\ relative to the PC population due to strong dilution from an additional star in the photometric aperture.  The transit model has to match both the transit duration and depth.  When dilution is present, a smaller transiting object and lower density (larger radius) star are fit to the observed transit. 

The third FP population is the remaining cloud of points, indicated by blue trianges, are the intrinsic EB population that have measured transit radii larger than 20 \rearth.  The transit value of the mean stellar density (\rhoc) is systematically different from the true value as our transit model is based on Equation \ref{rhostar} that assumes that the orbiting companion emits no light and has zero mass.


The PC population shows an overabundance of candidates at short durations that are offset towards lower mean stellar densities and may represent a population of unidentified BGEBs. An examination of PCs with a transit duration between 1 and 2 hours and \rhoc\ of $\sim$2 \gcmc\ shows a population of PCs with periods less than 2 days and PCs in multi-planet systems that were not validated in \citet{Lissauer2014} and \citet{Rowe2014}.  These PCs were not validated due to problems with centroid offsets.  There will also be a population of PCs that will have systematic errors in a comparison of \rhoc\ and \rhostar\ due to dilution from being members of hierarchical triples.

\subsection{A Transit HR Diagram}

The bottom panel of Figure \ref{hrdiag} plots the stellar \teff\ based on our adopted stellar properties in Table \ref{spars} \citep{Huber2014} vs \rhoc\ from our transit models.  It can be directly compared to the panel above based only on the stellar parameters.  Isochrones are based on the Dartmouth stellar evolution models \citep{Dotter2008} and plotted for \feh=-2.0,0.0,+0.5 and ages of 1 and 14 Gyr.  There is good agreement compared to the model isochrones.  Most of the  PCs that have \rhoc\ $<$ 1 \gcmc\ have \teff\ $>$ 5500 K, which is where one expects to see evolved stars with transiting planets.  More massive, hotter stars have relatively short main-sequence lifetimes.  The isochrones predict that only G and earlier type stars in the \ik FOV will have had time to show significant evolution off the main-sequence. The rest of the spread can be attributed to measurement error, metallicity of host-star, eccentric orbits and planetary systems associated with hierarchical triples.   

Measurement error tends to spread the determination of \rhoc\ evenly in both directions.  Eccentricity is biased towards larger values of \rhoc\ as there is a high probability of seeing a transiting planet near periastron.  When a planetary system is part of a hierarchical triple, there will be dilution from the extra star and as described in \S\ref{dissect} the value of \rhoc\ will be systematically lower.  Figure \ref{hrdiag} shows that all three effects are present and transit models can be used to measure the distribution and rate of eccentricity and hierarchical triples, see for example \citet{Rowe2014} or \citet{Moorhead2011}.

While a careful modeling of the effects of eccentricity and hierarchical triples is beyond the scope of this article, it is important to point out that there are a large number of PCs around cool host star (\teff\ $<$ 4000 K) that have smaller values of \rhoc\ than predicted by the overlaid isochrones.  This suggests that stellar binarity and dilution are important factors for M-stars and the radii of many PCs with cool host stars may be underestimated.  This is apparent in \S\ref{hzcandidates} on HZ candidates, where most of the host stars are cool relative to the Sun.

\subsection{Systems with multiple planet candidates}

Many new multiplanet systems are identified in this catalog, and many previously identified systems are either not identified or have been identified as false positives.  Here we give a brief overview of the new multiplanet systems and identify differences between this catalog and the catalog of \citep{Burke2014} (which used data through Quarter 10 to identify multi's).  For this comparison, we select all multiplanet systems in each catalog that do not have any planet pairs with a period ratio smaller than 1.1 (eliminating putative systems that are likely to be dynamically unstable or split multiplanet systems such as Kepler-132 (KOI-284).

\begin{figure}
\includegraphics[width=0.45\textwidth]{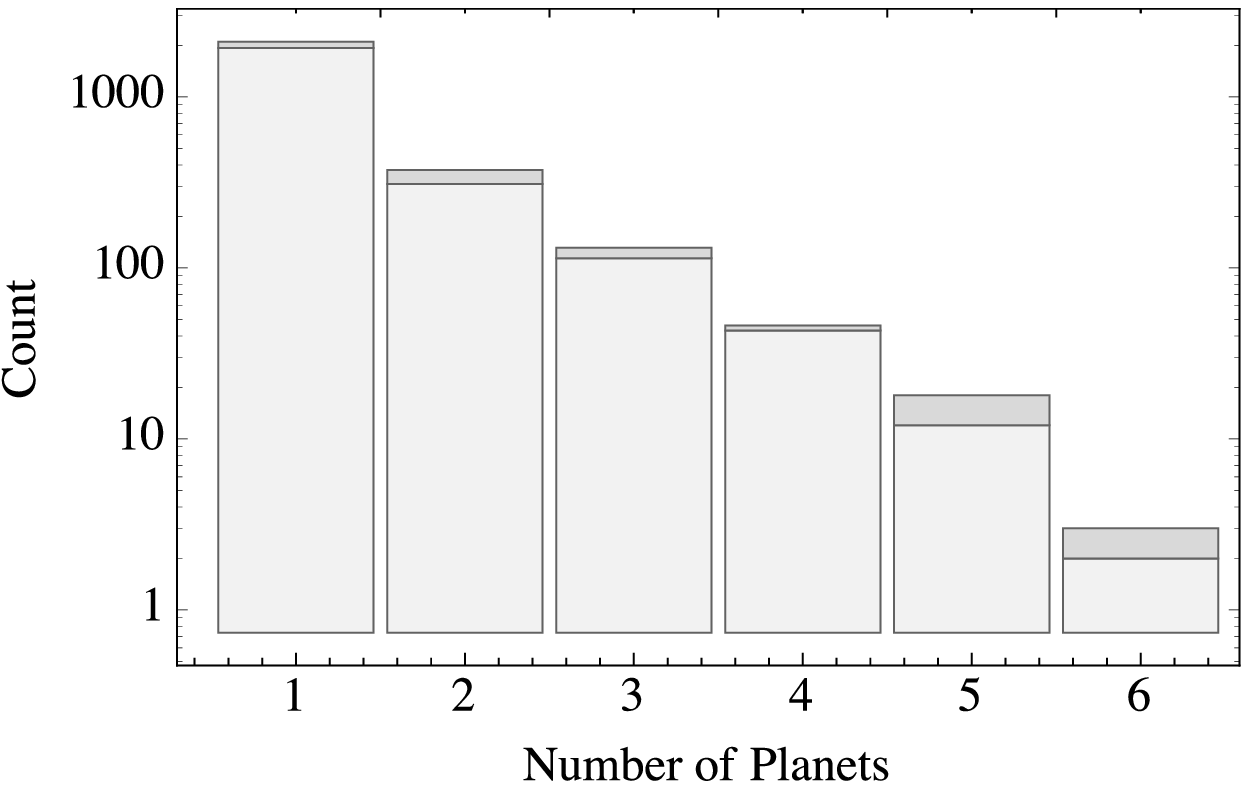}

\medskip
\includegraphics[width=0.45\textwidth]{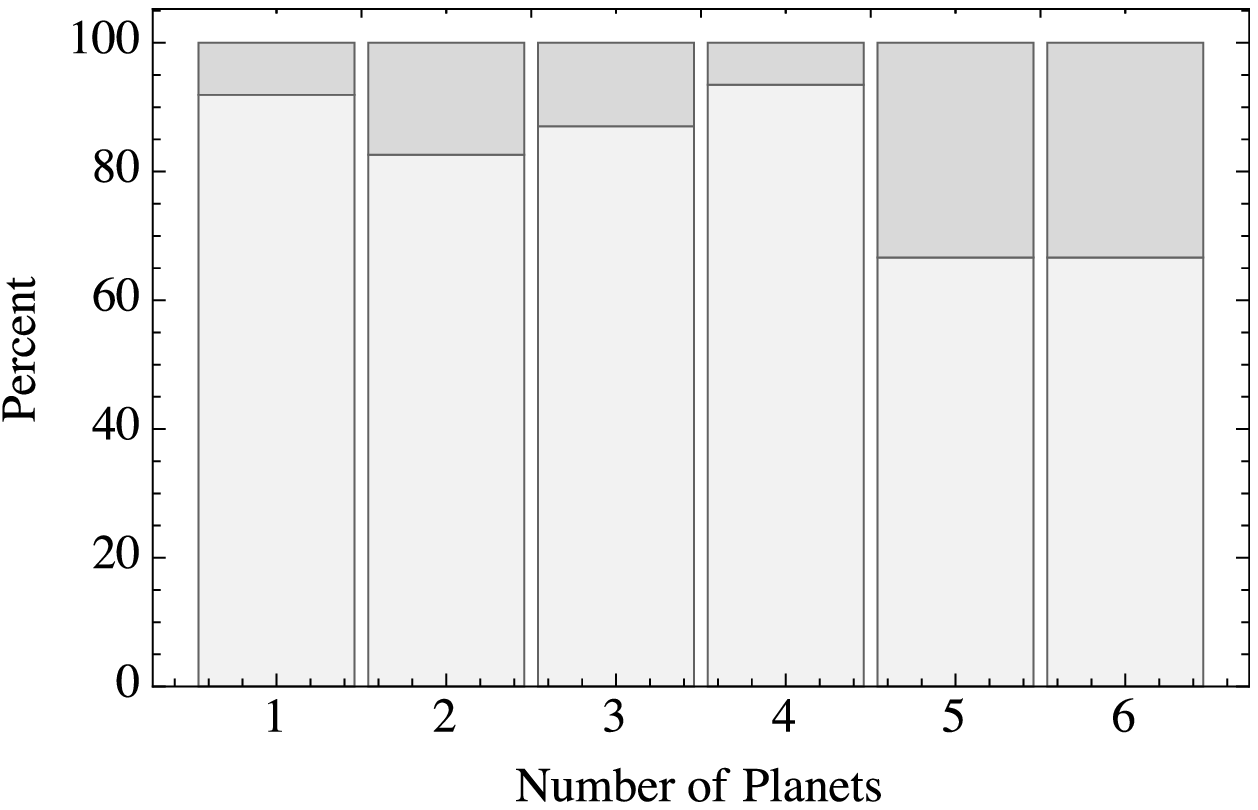}
\caption{The number of KOI systems versus system multiplicity.  Total from the \citet{Burke2014} catalog are the lighter gray and totals from this catalog are a darker gray.  The top panel shows raw counts and the bottom panel shows the contributions of the two catalogs to the total.\label{histograms}}
\end{figure}

The Q1-Q8 catalog has \oldsys\ unique KOI systems with \oldmulti\ of them multi-KOI systems.  The Q1-Q8 systems comprise \oldplans\ total planet candidates with \oldmultiplans\ of the candidates in multi-KOI systems.  The candidate yield in this new catalog increases to \newsys\ total KOI systems with \newmulti\ multi-KOI systems.  The new systems comprise \newplans\ total candidates with \newmultiplans\ of the total being in multi-KOI systems.  These gains in KOI yield are in spite of the loss of 400 KOIs from the Q1-Q8 catalogue \citep{Burke2014}, that are now identified as false positives.  The multiplicity of the 92 new multi-KOI systems include a net gain of 65 two-planet, 17 three-planet, 3 four-planet, 6 five-planet, and one six-planet system.  Figure \ref{histograms} shows a histogram of the system multiplicities from the previous and new multiplanet systems.

Of the KOI systems that are common to the Q1-Q8 and Q1-Q12 catalogs, many have different multiplicities.  There are \newlowkois\ new KOI systems in this catalog with KOI numbers less than 3149 (the largest numbered KOI in \citet{Burke2014}).  The balance of the new KOIs (\newhighkois ) are newly identified systems with numbers greater than 3149.

Among the common KOI systems, there are \uppers\ showing a net gain of planets---totaling \totaluppers\ new KOI.  Most of these changes are individual KOIs in a system, though KOI-2055 gained three candidates for a total of four and KOI-435 gained four candidates for a total of six.  At the same time, there are \downers\ where one or more candidates was not recovered in this pipeline.  All of the systems with changed multiplicities dropped only one candidate---KOI-5 is a notable example.

\begin{figure*}
\centering
\includegraphics[width=\linewidth]{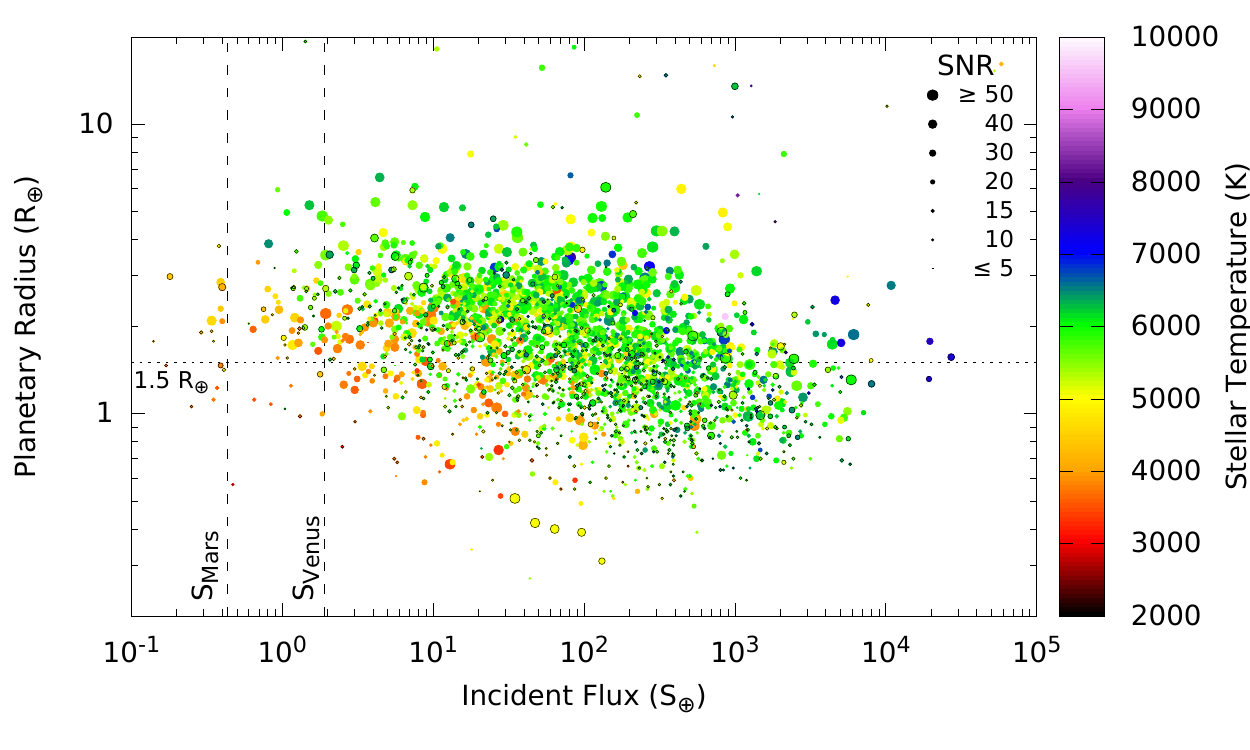}
\caption{A plot of planet radius versus incident flux for all planet candidates known in the Q1-Q12 catalogue. (Note that some planet candidates lie outside the chosen axis limits for the plot, and thus are not shown.) The temperature of the host star is indicated via the colour of each point, and the signal-to-noise of the detection is indicated via the size of each point. Planet candidates that were newly designated in Q1-Q12 are indicated with black circles around the point. The two vertical dashed lines indicate the incident flux recieved by Mars (0.43 $S_{\earth}$) and Venus (1.91 $S_{\earth}$), as a broad guide to a potential habitable zone. The horizontal dotted line is set at 1.5~$R_{\earth}$ as a suggested upper limit to terrestial-type planets.}
\label{planetparamfig}
\end{figure*}

\subsection{Early Type Stars}

Very little is known about the formation and evolution of planetary systems around hot stars. The small number of detected planets around stars hotter than \teff\ $>$ 6800 K is likely not to be intrinsic to the exoplanet population but rather the result of observational biases. Many early F and A-type stars are pulsating stars of type $\gamma$ Doradus and $\delta$ Scuti \citep[e.g.,][]{Uytterhoeven2011}. Their multi-periodic variability, with amplitudes up to several millimagnitudes, make it very complicated to detect transitting planets. Furthermore, these stars have larger radii resulting in a smaller area of light being blocked by the planet, and therefore produce relatively shallow transits, which are more difficult to detect. Nevertheless, several planets have been discovered around (pulsating) A-type stars such as Formalhaut \citep[e.g.,][]{Currie2012}, beta Pictoris \citep{Koen2003} and WASP 33 \citep{CollierCameron2010}. In the catalogue presented here there are 42 PCs with effective temperatures higher than 6800 K. From those 42, 3 PCs are $\delta$ Scuti stars, 5 are $\gamma$ Doradus variables (3 of those are most likely eclipsing binaries rather than PC) and 3 are so-called hybrid stars exhibiting $\delta$ Scuti and $\gamma$ Doradus variability simultaneously. 

\subsection{HZ Candidates}\label{hzcandidates}

\begin{figure}[h]
\centering
\begin{tabular}{c}
\includegraphics[width=\linewidth]{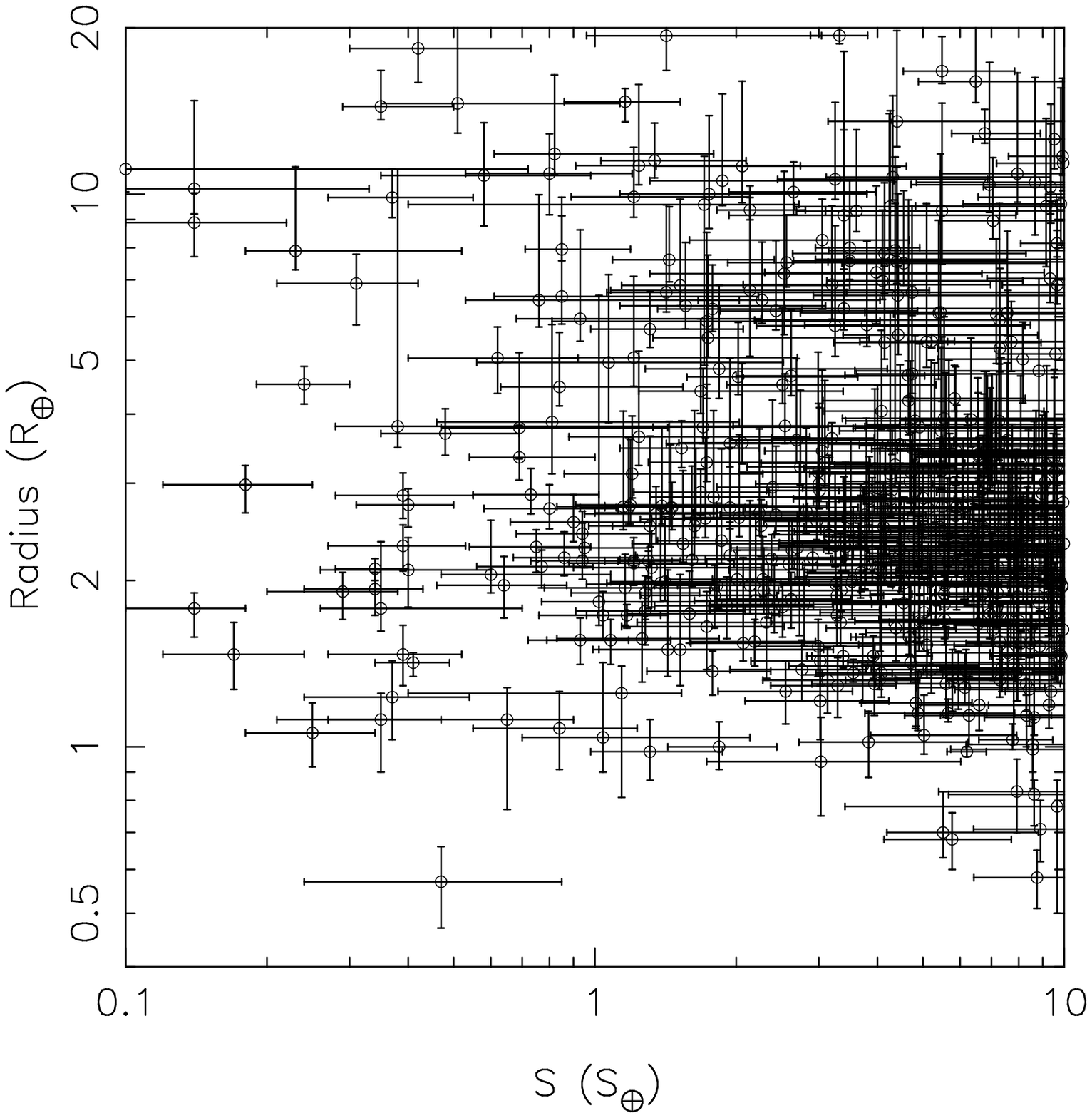}
\end{tabular}
\caption{Planet radius \rpl\ vs. incident flux ($S$) for PCs with $S<10$ $S_{\earth}$ and \rpl\ $<$ 20 \rearth.  Uncertainties are 1$\sigma$ based on posterior distributions calculated from Markov Chains based on transit models convolved with uncertainties from adopted stellar parameters.}
\label{HZplot}
\end{figure}

Figure \ref{planetparamfig} plots the Q1-Q12 PCs as a function of incident flux ($S$) vs \rpl\ with colours representing \teff\ of the host star and point sizes representing signal-to-noise.  As the transit search is based on 3 years of photometry our search was not sensitive to finding three transits of small Earth-sized planets in one year orbits around Sun-like stars, mostly due to stellar noise \citep{Gilliland2011}.  Such incompleteness is evident by noting that in Figure \ref{planetparamfig} there is an overabundance of small radius PCs in the HZ around cool (\teff\ $<$ 4000 K) stars.  Figure \ref{HZplot} shows a close up of PCs with 1$\sigma$ uncertainties based on our MCMC analysis and Table \ref{HZtable} lists 14 HZ PCs with \rpl\ $<$ $1.5\ \rearth$ and $S$ $<$ 2.  Kepler-62e \citep{Borucki2013} is not listed as its fitted radius is 1.73 \rearth.  KOI-4878.01 is a low S/N event.  As stated in \S \ref{snr}, for any KOI with a S/N $\lesssim$ 10 there is non-negligible probability that the transit event is not real, thus KOI-4878 should be treated with caution. 

Other than KOI-4878.01, which is likely a FA, all of the HZ candidates listed in Table \ref{HZtable} have cool K or M-dwarf host stars.  While M-stars are the most common star in the galaxy, these hosts present unique challenges towards potential habitability due to short orbital separation of the planet \citep{Tarter2007}, difficulty in accreting and reatining H$_2$O \citep{Lissauer2007} and phenomena such as stellar flares \citep{Segura2010}.  We examine each of the HZ candidates listed in Table \ref{HZtable} and give a brief discription of the characteristics of the system including the presence of strong stellar activity which, when present, presents evidence that the transiting HZ candidate is not a background blend.

KOI-3138.01 appears to be an interesting sub-Earth radius planet-candidate in a 8.7 day period around a cool M-dwarf (\teff = 2703 K).  The star was identified as a high proper motion target \citep{Lepine2005} (0.157"/yr). The fitted value of the mean-stellar density (\rhostar$_c$) of $70\pm^{25}_{42}$ \gcmc\ and short transit-duration also agree that the host star is compact, consistent with a late dwarf.  The object was added in Q6 as a Kepler-GO target to search for lensing and only long-cadence (30 minute) observations are available.  The star was unclassified in the Q1-Q8 catalogue, but its updated nominal properties make the planet Mars-sized (\rpl\ = 0.57 \rearth) that receives a Mars-like amount of flux ($S$ = 0.47 $S_{\earth}$).

KOI-3284.01 is an Earth-sized PC (\rpl = 0.98 \rearth) that receives 31\% more flux than the Earth and orbits a cool M-dwarf (\teff = 3688 K).  The photometric lightcurve shows 2\% variations consistent with spot modulation from a star with a spin period of 36 days.

KOI-2418.01 and KOI-2626.01 are modeled as two Earth-sized PCs (\rpl= 1.1 \rearth) that receive approximately one-third and two-thirds the flux the Earth receives and orbit stars classified as cool M-dwarfs with periods of 86 and 38 days, respectively, which have similar characteristics to Kepler-186f \citep{Quintana2014}.  The transit model values of \rhostar$_c$, $3.5\pm^{0.3}_{1.7}$ and $5.5\pm^{4.0}_{1.1} \gcmc$, are consistent with the stellar classification.   KOI-2418 shows relatively large (0.8\%) photometric variability due to star spots and a rotation period of 19 days and appears to show stellar flares. KOI-2626.01 has been observed to be an optical triple thus the planetary radius reported is underestimated \citep{Star2014}.  

KOI-2650.01 is part of a multi-planet candidate system.  This candidate has a orbital period of 34.99 days and \rpl\ = 1.25 \rearth .  The second candidate, KOI-2650.02, has an orbital period of 7.05 days, which produced a period ratio $P_{.01}/P_{.02} = 4.96$.  The high-order mean-resonance would would produce significant TTVs if the planets had high eccentricities.  There are no signs of TTVs for KOI-2650.01, and any potential TTVs KOI-2650.02 are not convincing.  We have not ruled out that any possible TTVs may be due to star spots.  The host star shows 2\% spot modulations consistent with a 20-day rotation period.  

KOI-2124.01 and 3255.01 are Earth-sized (\rpl\ = 1.0 and 1.4 \rearth), but receive $\sim$80\% more flux than the Earth, and thus these system have a stronger resemblance to Venus than the Earth \citep{Kane2014}.   KOI-2124 shows star spot modulation with photometric variability of 0.6\% with a 16-day period, there is also evidence of flares.  KOI-3255 shows variability greater than 1\%, consistent with star spots and a rotation period of 22 days.

Of the dozen credible HZ candidates presented, at least two are known binaries and thus, this highlights the importance of follow up of these systems with both spectroscopy and high resolution imaging \citep{Marcy2014,Gilliland2015}.

\section{Summary}

From an analysis of 18,406 TCEs we have added \nqtwelvenewpcs new PCs to the KOI database to bring the total number of PCs to \nallpcsk. \ik has now discovered more than a dozen good HZ candidates that have radii less than 1.5 \rearth\ and $S$ less than 2.0 $S_{\earth}$ primarily around cool dwarf stars.  We also deliver, for the first time, a uniform MCMC analysis of all KOI PCs and present reliable posterior distributions convolved with improved stellar classifications of \ikt's target stars.  Our transit curve analysis is extremely useful, not only to determine fundamental properties of extrasolar planets, but to also cull the population of KOIs to select a highly reliable set of planet candidates based on period, S/N, transit duration and depth. With more than 4 quarters of \ik photometry left to analyze, and still improving data analysis software, we are excited about the future prospects of \ik discoveries.

\begin{deluxetable}{ccccc}
\tabletypesize{\scriptsize}
\tablecaption{Small HZ Planets and Candidates}
\tablehead{\colhead{KOI} & \colhead{\teff} & \colhead{\rpl} & \colhead{$S$} & \colhead{S/N} \\ 
\colhead{} & \colhead{K} & \colhead{\rearth} & \colhead{$S_{\earth}$} & \colhead{}}
\startdata
571.05$^1$ & 3761 & 1.06 & 0.25 & 12.4 \\
701.04$^2$ & 4797 & 1.42 & 0.41 & 18.1 \\
1422.04$^3$ & 3517 & 1.23$^4$ & 0.37 & 17.0 \\
1422.05$^3$ & 3517 & 1.08$^4$ & 0.84 & 14.0 \\
2124.01 & 4029 & 1.00 & 1.84 & 21.6 \\
2418.01 & 3724 & 1.12 & 0.35 & 16.7 \\
2626.01 & 3482 & 1.12$^4$ & 0.65 & 16.2 \\
2650.01 & 3735 & 1.25 & 1.14 & 14.1 \\
3138.01 & 2703 & 0.57 & 0.47 & 10.8 \\
3255.01 & 4427 & 1.37 & 1.78 & 27.0 \\
3284.01 & 3688 & 0.98 & 1.31 & 16.4 \\
4087.01 & 3813 & 1.47 & 0.39 & 23.9 \\
4427.01 & 3668 & 1.47 & 0.17 & 13.7 \\
\enddata
\tablecomments{List of potential HZ candidates with \rpl $<$ 1.5 \rearth, $S$ $<$ 2 $S_{\earth}$ and S/N $>$ 10.  Any candidate with a S/N less than $\sim$10 should be considered unreliable\\
Notes: $^1$ Kepler-186f, \\
$^2$ Kepler-62e, \\
$^3$ Kepler-296e, Kepler-296f, \\
$^4$ known binary, thus \rpl is underestimated. \\
}
\label{HZtable}
\end{deluxetable}

\acknowledgments

Funding for this Discovery mission is provided by NASA's Science Mission Directorate.  We are grateful to TCERT vetters who tirelessly examined thousands of transit candidates. We are indebted to the entire \ik Team for all the hard work and dedication that have made such discoveries possible.  In one way or another, it seems that everyone in the exoplanet community has somehow contributed towards this work and if I add everyone to the author list there will no one left to referee, so thank you everyone and the referee.  J.F.R. acknowledges NASA grants NNX12AD21G and NNX14AB82G issued through the Kepler Participating Scientist Program.  B.Q. acknowledges support from a NASA Postdoctoral Fellowship.  D.H. acknowledges NASA Grant NNX14AB92G issued through the Kepler Participating Scientist Program and support by the Australian Research Council's Discovery Projects funding scheme (project number DE140101364).  Funding for the Stellar Astrophysics Centre is provided by The Danish National Research Foundation (grant No. DNRF106). V.A. is supported by the ASTERISK project (ASTERoseismic Investigations with SONG and Kepler) funded by the European Research Council (Grant agreement No. 267864).  K.G.H. acknowledges support provided by the National Astronomical Observatory of Japan as Subaru Astronomical Research Fellow. This research has made use of the NASA Exoplanet Archive, which is operated by the California Institute of Technology, under contract with the National Aeronautics and Space Administration under the Exoplanet Exploration Program.

\bibliography{AstroRefs.bib}

\newpage
\begin{turnpage}
\tabcolsep=0.11cm
\begin{deluxetable}{ccccccccccccc}
\tabletypesize{\scriptsize}
\tablecaption{Transit Model Parameters}
\tablewidth{0pt}
\tablehead{
\colhead{KOI}  & \colhead{KIC} & \colhead{$P$}  & \colhead{\rpl}    & \colhead{$S$}              & \colhead{$b$} & \colhead{\rprs} & \colhead{\rhostar$_{f}$} & \colhead{T$_{dep}$} & \colhead{T$_{dur}$} & \colhead{$T_0$}        & \colhead{S/N} & \colhead{fp} \\
\colhead{}     & \colhead{}          & \colhead{days} & \colhead{\rearth} & \colhead{$S_{\earth}$}     &  \colhead{}   & \colhead{}      & \colhead{\gcmc}          & \colhead{ppm}       & \colhead{hours}     & \colhead{days$^1$}  & \colhead{}    & \colhead{} 
}
\startdata
1.01 & 11446443 & 2.4706134 & 12.850 & 772.2 & 0.82 & 0.12385 & 1.8318 & 14186.4 & 1.743 & 55.763301 & 6802.0 & 0 \\
  &   & 0.0000000 & +0.270-0.270 & +60.7-57.1 & +0.00-0.00 & +0.00003-0.00008 & +0.0068-0.0044 & 46.7 & 0.001 & 0.000006 &   &   \\
2.01 & 10666592 & 2.2047354 & 16.390 & 3973.7 & 0.00 & 0.07541 & 0.4059 & 6690.6 & 3.882 & 54.358572 & 6714.5 & 0 \\
  &   & 0.0000000 & +0.150-0.140 & +279.9-264.5 & +0.01-0.00 & +0.00001-0.00001 & +0.0001-0.0003 & 1.3 & 0.000 & 0.000014 &   &   \\
3.01 & 10748390 & 4.8878027 & 4.840 & 97.1 & 0.03 & 0.05799 & 3.7001 & 4342.1 & 2.364 & 57.813141 & 2207.8 & 0 \\
  &   & 0.0000002 & +0.190-0.140 & +16.2-12.3 & +0.05-0.03 & +0.00005-0.00003 & +0.0114-0.0291 & 2.2 & 0.001 & 0.000028 &   &   \\
4.01 & 3861595 & 3.8493715 & 13.100 & 4055.3 & 0.92 & 0.04011 & 0.2082 & 1317.3 & 2.661 & 90.526738 & 262.6 & 1 \\
  &   & 0.0000013 & +2.060-3.250 & +1837.2-1919.9 & +0.01-0.01 & +0.00028-0.00047 & +0.0250-0.0202 & 7.3 & 0.034 & 0.000269 &   &   \\
5.01 & 8554498 & 4.7803278 & 7.070 & 898.7 & 0.95 & 0.03707 & 0.3442 & 977.2 & 2.035 & 65.974137 & 383.4 & 0 \\
  &   & 0.0000009 & +0.170-0.170 & +93.8-86.5 & +0.00-0.00 & +0.00016-0.00023 & +0.0175-0.0135 & 4.0 & 0.014 & 0.000152 &   &   \\
5.02 & 8554498 & 7.0518600 & 0.200 & 534.8 & 0.95 & 0.00104 & 0.3479 & 0.8 & 1.740 & 66.367000 & 0.3 & 1 \\
  &   & -- & -- & -- & -- & -- & -- & -- & -- & -- &   &   \\
6.01 & 3248033 & 1.3341043 & 50.730 & 5207.6 & 1.27 & 0.29402 & 0.0353 & 444.2 & 3.014 & 66.701635 & 192.7 & 1 \\
  &   & 0.0000007 & +13.320-10.920 & +3668.2-2267.5 & +0.10-0.22 & +0.10368-0.20946 & +0.0070-0.0094 & 3.1 & 0.022 & 0.000420 &   &   \\
7.01 & 11853905 & 3.2136686 & 4.140 & 1218.9 & 0.02 & 0.02474 & 0.4638 & 727.7 & 3.994 & 56.611934 & 328.5 & 0 \\
  &   & 0.0000011 & +0.110-0.110 & +133.3-122.4 & +0.20-0.02 & +0.00014-0.00008 & +0.0002-0.0571 & 2.6 & 0.009 & 0.000280 &   &   \\
8.01 & 5903312 & 1.1601532 & 2.000 & 2229.3 & 0.78 & 0.01856 & 1.0186 & 368.7 & 1.413 & 54.704057 & 201.0 & 1 \\
  &   & 0.0000004 & +0.380-0.160 & +1175.9-491.1 & +0.02-0.56 & +0.00025-0.00168 & +2.5637-0.1790 & 19.3 & 0.025 & 0.000369 &   &   \\
9.01 & 11553706 & 3.7198080 & 7.850 & 616.1 & 0.94 & 0.07082 & 0.1069 & 3749.1 & 3.522 & 68.068333 & 587.5 & 1 \\
  &   & 0.0000007 & +3.600-0.700 & +825.4-162.4 & +0.00-0.00 & +0.00090-0.00093 & +0.0022-0.0035 & 15.8 & 0.014 & 0.000168 &   &   \\
10.01 & 6922244 & 3.5224986 & 14.830 & 1264.7 & 0.61 & 0.09358 & 0.6891 & 9379.3 & 3.191 & 54.119429 & 1801.5 & 0 \\
  &   & 0.0000002 & +1.190-1.320 & +323.4-317.5 & +0.01-0.01 & +0.00012-0.00020 & +0.0184-0.0125 & 6.6 & 0.006 & 0.000046 &   &   \\
11.01 & 11913073 & 3.7478392 & 10.470 & 266.8 & 1.08 & 0.11602 & 0.0108 & 871.7 & 5.111 & 104.664884 & 200.6 & 1 \\
  &   & 0.0000032 & +3.870-0.750 & +293.0-59.1 & +0.14-0.07 & +0.13034-0.06166 & +0.0008-0.0007 & 5.8 & 0.036 & 0.000723 &   &   \\
12.01 & 5812701 & 17.8552197 & 14.630 & 186.2 & 0.00 & 0.08839 & 0.4742 & 9153.6 & 7.429 & 79.596388 & 1034.5 & 0 \\
  &   & 0.0000038 & +8.540-3.180 & +322.2-86.3 & +0.05-0.00 & +0.00008-0.00005 & +0.0004-0.0042 & 11.4 & 0.005 & 0.000170 &   &   \\
13.01 & 9941662 & 1.7635876 & 25.800 & 37958.3 & 0.37 & 0.07794 & 0.4963 & 4598.9 & 3.181 & 53.565925 & 5120.6 & 0 \\
  &   & 0.0000000 & +10.190-8.040 & +44636.8-23093.5 & +0.01-0.01 & +0.00005-0.00004 & +0.0044-0.0065 & 1.9 & 0.005 & 0.000015 &   &   \\
14.01 & 7684873 & 2.9473757 & 5.890 & 7903.8 & 0.98 & 0.02414 & 0.1103 & 401.0 & 1.966 & 104.523181 & 75.6 & 1 \\
  &   & 0.0000006 & +2.730-1.310 & +11190.2-3801.2 & +0.00-0.00 & +0.00022-0.00030 & +0.0039-0.0067 & 3.0 & 0.016 & 0.000171 &   &   \\
15.01 & 3964562 & 3.0124768 & 92.010 & 6499.9 & 1.41 & 0.44868 & 0.0521 & 1861.7 & 3.095 & 68.259011 & 345.1 & 1 \\
  &   & 0.0000019 & +65.660-13.130 & +14930.2-2510.4 & +0.30-0.27 & +0.28730-0.26741 & +0.0135-0.0081 & 16.1 & 0.027 & 0.000518 &   &  
\enddata
\tablecomments{fp: 0 - planet-candidate, 1 - False-positive, $^1$T0=BJD-2454900}
\label{mplanetfit}
\end{deluxetable}

\end{turnpage}

\clearpage

\appendix
\label{appendixsec}

\section{List of symbols and abbreviations}

\begin{itemize}
\item BGEB -- background eclipsing binary
\item DV -- Data Validation
\item EB -- eclipsing binary
\item FA -- False-alarms
\item FP -- False-positive
\item HZ - habitable zone
\item KOI -- Kepler Object of Interest
\item \mstar, \rstar\ -- mass and radius in solar units
\item MCMC -- Markov Chain Monte Carlo
\item MES -- Multiple Event Statistic
\item OOT -- out-of-transit
\item PC -- planetary candidate
\item PDC -- pre-search data conditioning
\item PRF -- point response function 
\item \rearth\ -- radius relative to the Earth
\item \rhoc\ -- transit model derived mean stellar density for circular orbits
\item \rhostar\ -- mean stellar density
\item $S$ -- ratio of incident flux relative to the Earth.
\item S/N -- signal-to-noise ratio.
\item TCE -- Threshold Crossing Event
\item TCERT -- Threshold Crossing Event Review Team
\item TTV -- transit timing varation
\end{itemize}

\begin{figure*}
\centering
\includegraphics[width=\textwidth,height=\textheight,keepaspectratio]{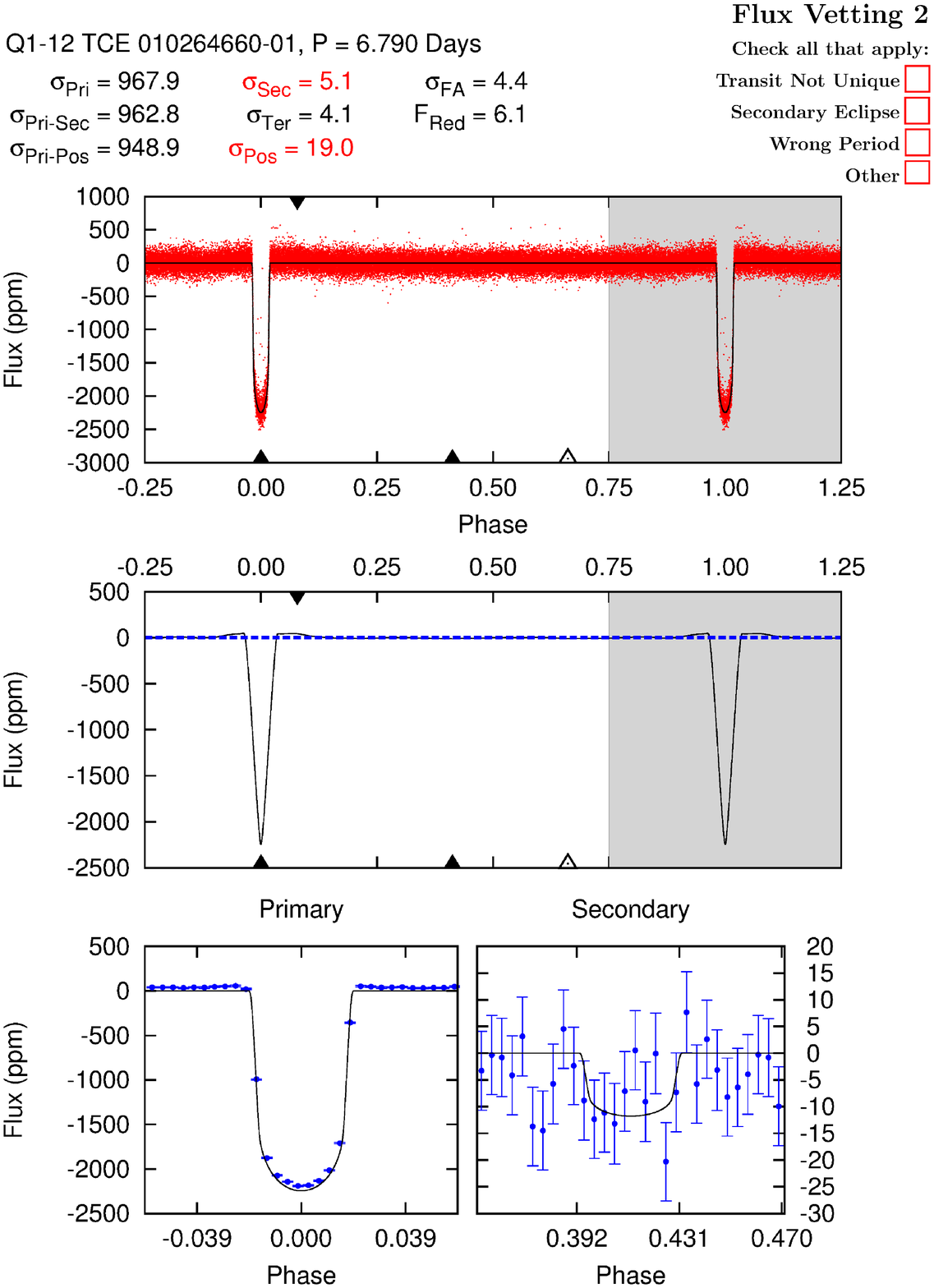}
\caption{Page 2 of the Q1-Q12 TCERT Dispositioning form for Kepler-14b, a well-known confirmed planet.}
\label{disppdffig2}
\end{figure*}

\begin{figure*}
\centering
\includegraphics[width=\textwidth,height=\textheight,keepaspectratio]{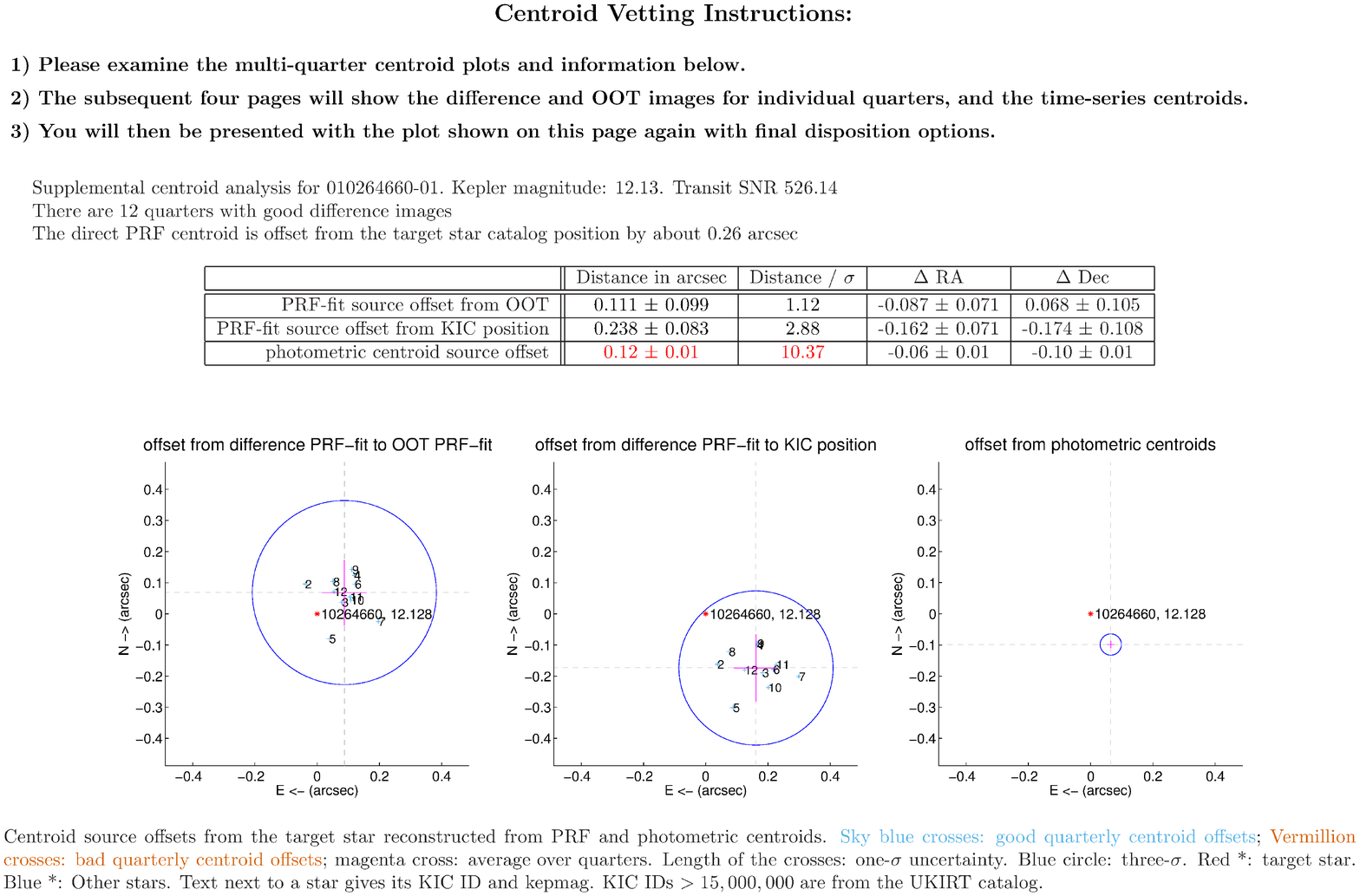}
\caption{Page 3 of the Q1-Q12 TCERT Dispositioning form for Kepler-14b, a well-known confirmed planet.}
\label{disppdffig3}
\end{figure*}

\begin{figure*}
\centering
\includegraphics[width=\textwidth,height=\textheight,keepaspectratio]{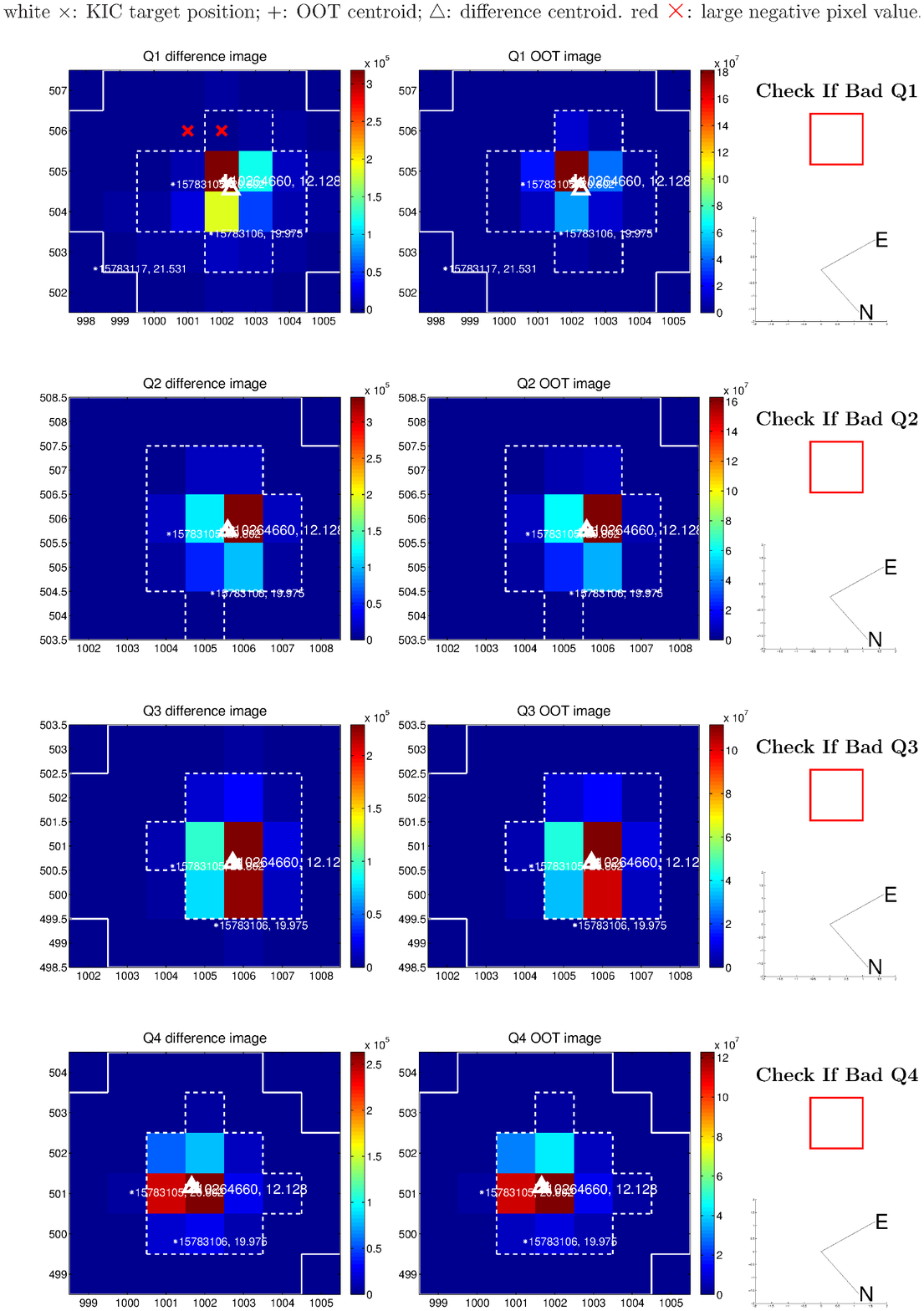}
\caption{Page 4 of the Q1-Q12 TCERT Dispositioning form for Kepler-14b, a well-known confirmed planet.}
\label{disppdffig4}
\end{figure*}

\begin{figure*}
\centering
\includegraphics[width=\textwidth,height=\textheight,keepaspectratio]{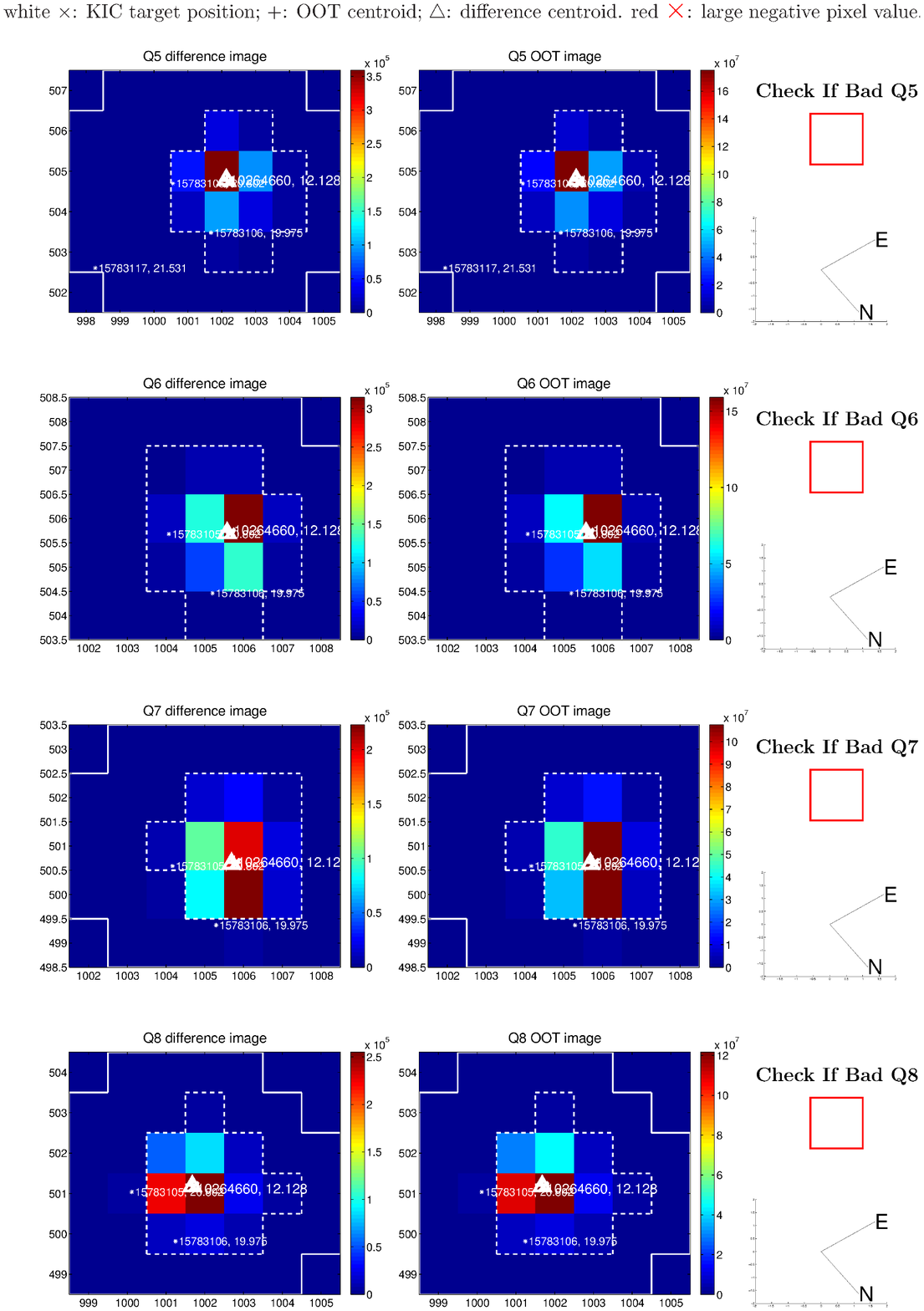}
\caption{Page 5 of the Q1-Q12 TCERT Dispositioning form for Kepler-14b, a well-known confirmed planet.}
\label{disppdffig5}
\end{figure*}

\begin{figure*}
\centering
\includegraphics[width=\textwidth,height=\textheight,keepaspectratio]{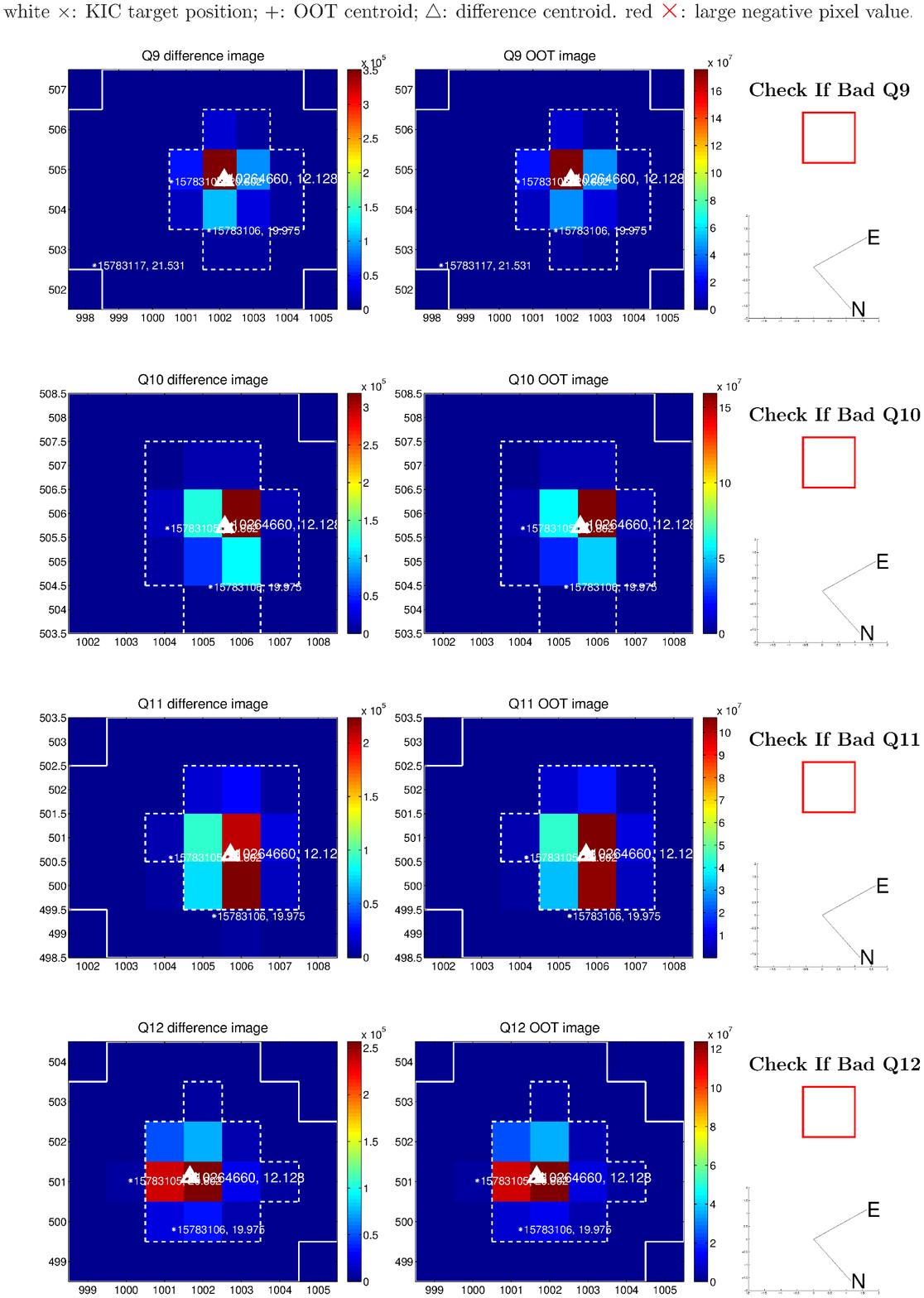}
\caption{Page 6 of the Q1-Q12 TCERT Dispositioning form for Kepler-14b, a well-known confirmed planet.}
\label{disppdffig6}
\end{figure*}

\begin{figure*}
\centering
\includegraphics[width=\textwidth,height=\textheight,keepaspectratio]{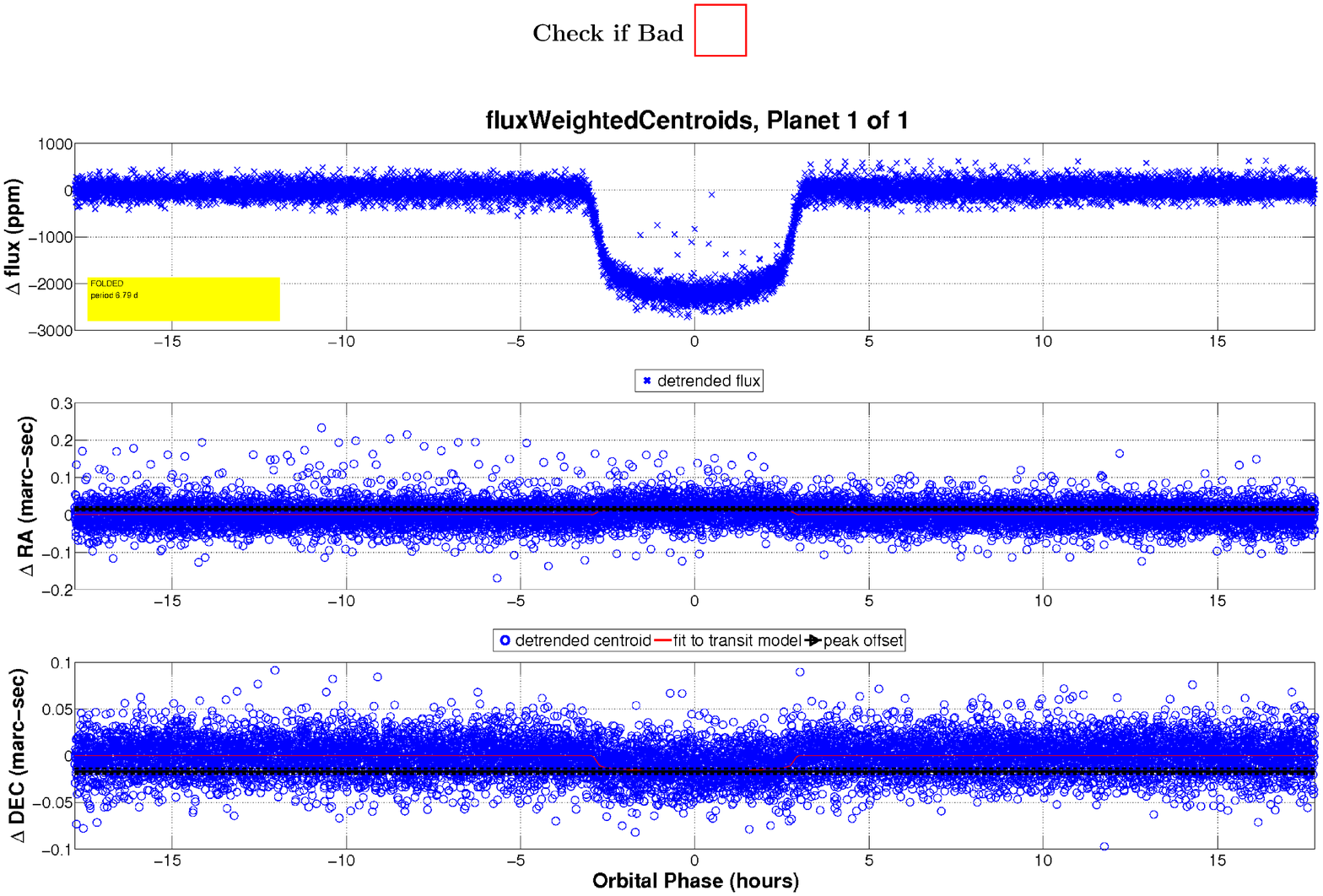}
\caption{Page 7 of the Q1-Q12 TCERT Dispositioning form for Kepler-14b, a well-known confirmed planet.}
\label{disppdffig7}
\end{figure*}

\begin{figure*}
\centering
\includegraphics[width=\textwidth,height=\textheight,keepaspectratio]{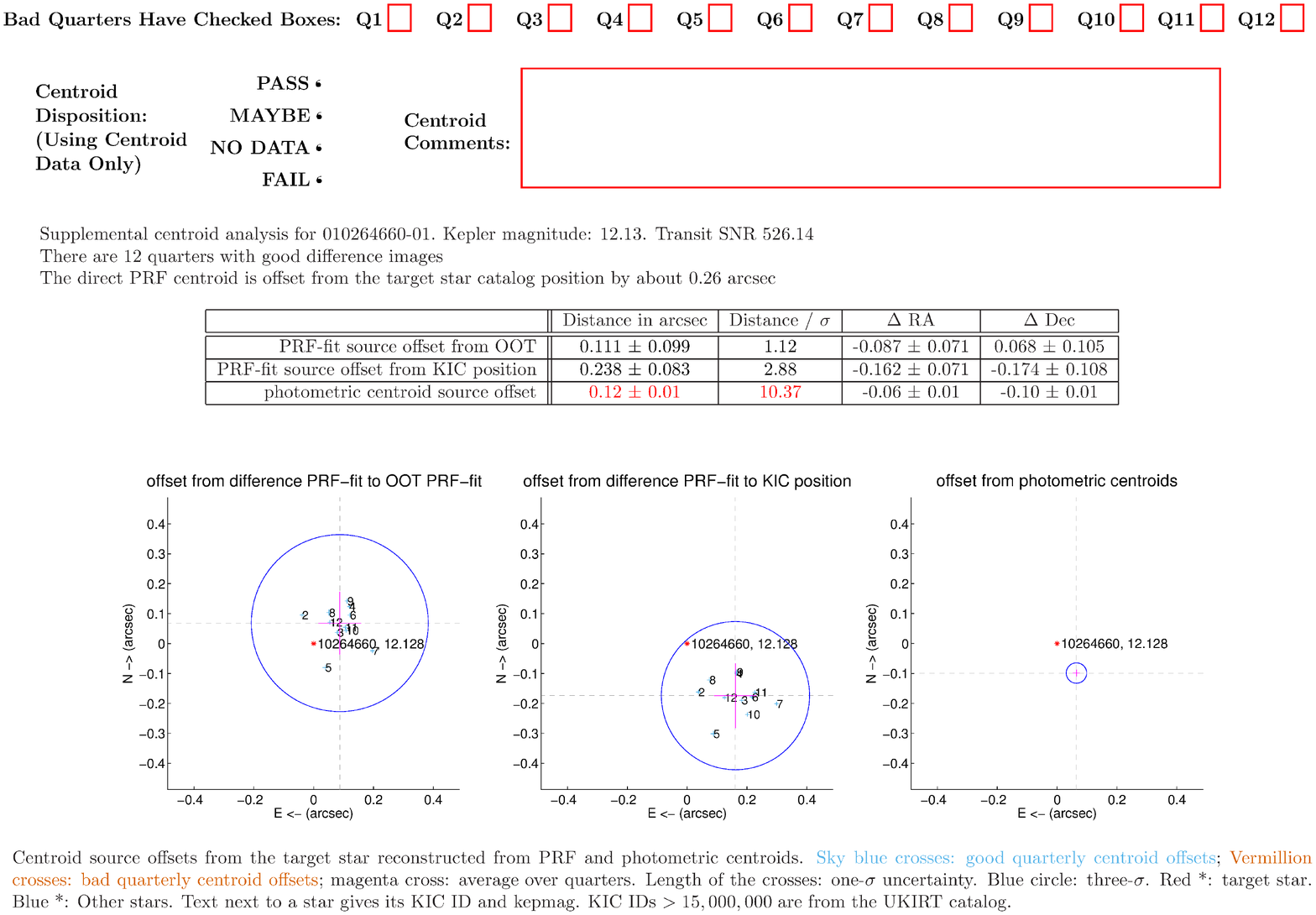}
\caption{Page 8 of the Q1-Q12 TCERT Dispositioning form for Kepler-14b, a well-known confirmed planet.}
\label{disppdffig8}
\end{figure*}

\end{document}